\documentstyle [12pt,amsfonts] {article}
\input epsf
\newcommand{\beq}{\begin{equation}}
\newcommand{\eeq}{\end{equation}}
\newcommand{\beqs}{\begin{eqnarray}}
\newcommand{\eeqs}{\end{eqnarray}}

\newcommand{\gtwid}{\mathrel{\raise.3ex\hbox{$>$\kern-.75em\lower1ex
\hbox{$\sim$}}}}
\newcommand{\ltwid}{\mathrel{\raise.3ex\hbox{$<$\kern-.75em\lower1ex
\hbox{$\sim$}}}}
\newcommand{\bi}{\bibitem}

\catcode`@=11
\@addtoreset{equation}{section}
\@addtoreset{equation}{subsection}
\def\theequation{\ifnum\value{section}=0 \arabic{equation}\ignorespaces
\else \ifnum\value{section}=-1 A.\arabic{equation}\ignorespaces
\else \ifnum\value{subsection}=0
\thesection.\arabic{equation}\ignorespaces
\else \thesection.\arabic{subsection}.\arabic{equation}\ignorespaces
                           \fi
                      \fi
                 \fi}
\catcode`@=12

\newtheorem{th}{Theorem}[section]

\begin{document}

\begin{titlepage}

\begin{center}

{\Large \bf Reliability Polynomials and their Asymptotic Limits for Families of
Graphs}

\vspace{1.1cm}
{\large Shu-Chiuan Chang${}$\footnote{address after Sept. 2002:
Department of Applied Physics, Faculty of Science, Tokyo University of 
Science, Tokyo 162-8601, Japan; email: chang@rs.kagu.tus.ac.jp}
and Robert Shrock${}$\footnote{email:
robert.shrock@sunysb.edu }}\\
\vspace{18pt}
 C. N. Yang Institute for Theoretical Physics \\
 State University of New York \\
 Stony Brook, NY 11794-3840 \\
\end{center}

\vskip 0.6 cm

\begin{abstract}
\vspace{2cm}

We present exact calculations of reliability polynomials $R(G,p)$ for lattice
strips $G$ of fixed widths $L_y \le 4$ and arbitrarily great length $L_x$ with
various boundary conditions.  We introduce the notion of a reliability per
vertex, $r(\{G\},p) = \lim_{|V| \to \infty} R(G,p)^{1/|V|}$ where $|V|$ denotes
the number of vertices in $G$ and $\{G\}$ denotes the formal limit $\lim_{|V|
\to \infty} G$.  We calculate this exactly for various families of graphs.  We
also study the zeros of $R(G,p)$ in the complex $p$ plane and determine exactly
the asymptotic accumulation set of these zeros ${\cal B}$, across which
$r(\{G\})$ is nonanalytic.

\end{abstract}

\begin{flushleft}
keywords: reliability polynomial, Potts model, Tutte polynomial
\end{flushleft}

\end{titlepage}

\section{Introduction}

With the development of progressively larger communications networks such as
those involving telecommunications and those connecting computers, from large
local area networks to the internet, the analysis of the reliability of these
networks has become an increasingly important area of study.  Because many
types of failures of communication links are random, this study falls within
the area of statistical physics, as well as engineering and, as will be
evident, mathematical graph theory.  A common approach to this study is to
factorize the reliability into a factor describing the components such as
message routers and computers at various nodes, on the one hand, and the links
between these nodes, on the other hand. One often assumes that nodal network
elements like message routers and computers have a certain probability
$p_{node}$ of (normal) operation.  Since failures of these nodal components
generally occur independently, the factor in the overall reliability due to
these elements is then $p_{node}^n$, where $n$ is the number of nodes in
the network.  The other factor contributing to the overall network reliability
involves the connecting links on the network and hence the connectivity
structure of the network.  As mentioned above, many failures in these
connecting links are also random and independent and may thus be treated in a
probabilistic manner in which one describes a given link as having a
probability $p$ of operating and thus a probability $1-p$ of failing.
(Examples of nonrandom failures include earthquakes that affect only one
portion of a network; these are not considered here.)  The network is thus
represented as a connected graph $G=(V,E)$ with vertex (= node) set $V$ and
edge (= bond) set $E$, in which each vertex represents either a message router
or switch or an endpoint such as a computer or terminal (or telephone in the
telecommunications example) and each edge represents the link between the
nodes.  We shall be interested here in a commonly used simplified model of the
network in which each nodal element and link have, respectively, the same fixed
probabilities $p_{node}$ and $p$ of operation with $p_{node}$ and $p$ lying in
the interval $[0,1]$. We shall denote the number of vertices and edges in $G$
as $|V|=n$ and $|E|$.  In the graphical representation of the network, one
regards a normally operating communications link as an edge that is present
(with probability $p$), and a malfunctioning communications link as an edge
that is absent (with probability $1-p$), the presence or absence being random
and uncorrelated.  Defining a successfully operating network as one in which
all of links and nodes are operating normally, it follows that the node
reliability is, as noted above, $p_{node}^{|V|}$, independent of the
connectivity of the network. The factor due to the operation of the links is
much more difficult to calculate.  It is therefore customary, in modelling the
reliability behavior of networks, to separate out the node-reliability factor
and concentrate on the contribution due to the structure of the network.

The all-terminal reliability polynomial $R(G,p)$ is defined as the probability
that there is an operating communications link between any two nodes in the
network, i.e. any two vertices in the set $V$ are connected (by a path
consisting of edges that are present) \cite{shannon}-\cite{colbourn}.  In
passing, we note that one can also study $k$-terminal reliability probability
polynomials, but we shall restrict ourselves here to the all-terminal
reliability polynomial and hence shall henceforth omit the qualifier
``all-terminal'' in the notation.  The contributions to $R(G,p)$ arise from the
sum of connected spanning subgraphs of $G$. Here a spanning subgraph is a
subgraph $G^\prime=(V,E^\prime)$ with $E^\prime \subseteq E$, i.e. a graph with
the same vertex set and a subset of the edge set of $G$.  Denote a connected
spanning subgraph as $H=(V,E_H)$.  Each contribution to $R(G,p)$ is weighted by
a factor $p^{|E_H|}$ reflecting the probability that the $|E_H|$ edges are
present, multiplied by a factor $(1-p)^{|E|-|E_H|}$ reflecting the probability
that the edges in the complement set $\{e \in E; e \not\in E_H\}$, of 
cardinality $|E|-|E_H|$, are absent. That is,
\beq
R(G,p)=\sum_{H \subseteq G} p^{|E_H|}(1-p)^{|E|-|E_H|} \ .
\label{R}
\eeq
Some pioneering studies of network reliability include
\cite{shannon}-\cite{kelmans}; a review is \cite{colbourn}.  The calculation of
$R(G,p)$ for an arbitrary (connected) graph $G$ has been proved to be \#-P
complete \cite{colbourn,pb}; that is, roughly speaking, for a generic network
$G$, the time required to calculate $R(G,p)$ grows exponentially with the size
of the network.  In the face of this difficulty, network designers have often
relied upon upper and lower bounds on $R(G,p)$ \cite{lp}- \cite{bc96}. It is
clearly of interest if one can obtain exact solutions for $R(G,p)$ for some
families of graphs $G$.

In this paper, we shall present a number of such exact solutions for
reliability polynomials, for recursive families of graphs $G$.  We have
previously presented such solutions for a specific recursive family of graphs
in \cite{ka3}. A recursive family of graphs is one in which one constructs
successive members of the family in a recursive manner starting from an initial
member. Recursive families of graphs that are of particular interest here are
strips of regular lattices of a given width $L_y$ vertices and arbitrarily
great length $L_x$ vertices with some specified transverse and longitudinal
boundary conditions. We shall envision these strips as extending in the
horizontal (= longitudinal, $x$) direction and having transverse extent in the
vertical, $y$, direction.  To see that these form recursive families, one can
picture a strip, say of the square lattice, of length $L_x$ and width $L_y$ as
being formed by gluing on a column of squares of height $L_y$ to the strip of
length $L_x-1$.  The boundary conditions that we shall consider include free
(FBC), periodic (PBC), and twisted periodic (TPBC), by which is meant that the
longitudinal ends are identified with a reversal of orientation.  We shall
denote the various combinations as (i) $(FBC_y,FBC_x)=$ free, (ii)
$(FBC_y,PBC_x)=$ cyclic, (iii) $(FBC_y,TPBC_x)=$ M\"obius, (iv)
$(PBC_y,FBC_x)=$ cylindrical, (v) $(PBC_y,PBC_x)=$ torus, and (vi)
$(PBC_y,TPBC_x)=$ Klein bottle.  For an arbitrary graph $G$, one defines the
degree $d_i$ of a vertex $v_i$ as the number of edges connected to it.  One
then defines the maximum degree as $\Delta = \max_{v_i \in V} d_i$. A
$\Delta$-regular graph is one in which all of the vertices have the same degree
$\Delta$. Some strip graphs, such as those with toroidal or Klein bottle
boundary conditions, are $\Delta$-regular graphs.  For strip graphs that are
not $\Delta$-regular, it will still be useful to define an average or effective
vertex degree in the limit of infinite length, namely
\beq
d_{eff} = \lim_{|V| \to \infty} \frac{2|E|}{|V|} \ .
\label{delta_eff}
\eeq
Secondly, we shall introduce a definition of the reliability per node in the
limit of many nodes,
\beq
r(\{G\},p) = \lim_{|V| \to \infty} R(G,p)^{1/|V|} 
\label{r}
\eeq
where $\{G\}$ denotes the formal limit of a given family, $\lim_{|V| \to
\infty} G$.  We calculate this function exactly for a number of families and
study its properties.  As we shall show, this function provides a convenient
quantitative measure of the reliability properties of large networks.

A third part of our work involves the calculation of the zeros of $R(G,p)$ in
the complex $p$ plane and the determination of their asymptotic continuous
accumulation set, denoted ${\cal B}$ as the number of vertices $|V| \to \infty$
(in practice, for our strips, this is equivalent to the limit of infinite
length, $L_x \to \infty$).  As we shall show, the function $r(\{G\})$ is
nonanalytic across the locus ${\cal B}$.  We define $p_c$ as the maximal point
where ${\cal B}$ intersects (or coincides with) the real axis.  For the
infinite-length limits of some lattice strips, the locus ${\cal B}$ does not
cross the real axis, so no $p_c$ is defined.  In certain of these cases, there
are complex-conjugate arcs on ${\cal B}$ whose endpoints at $p_{end},
p_{end}^*$ are very close to the real axis; in these cases, it is useful to
define a $(p_c)_{eff} = Re(p_{end})$.  Although the width $L_y \to \infty$
limit of loci ${\cal B}$ for the infinite-length lattice strips is not
necessarily the same as the continuous accumulation locus of the zeros of the
reliability polynomial for the two-dimensional thermodynamic limit $L_x \to
\infty$, $L_y \to \infty$ with $L_y/L_x$ fixed to a finite nonzero constant,
our exact results on infinite-length limits of lattice strip graphs can give 
some insight into plausible behavior of the locus ${\cal B}$ for the
thermodynamic limits of the corresponding 2D lattices.  We shall discuss this
further below.

There are several motivations for our study.  Clearly, new exact calculations
of reliability polynomials are of interest from the point of view of both the
statistical physics of network theory and of mathematical graph theory.  The
generalization of $p$ from the interval $p \in [0,1]$ to $p \in {\mathbb C}$ is
clearly necessary in order to analyze the zeros of the reliability polynomial
and their accumulation set as $|V| \to \infty$.  Our results also further
illuminate the fascinating variety of applications of the Tutte polynomial, or
equivalently, the Potts model partition function.  Our focus here is thus
somewhat more abstract than in a specific engineering context, where one is
trying to design the most reliable network subject to various constraints, such
as cost.

\section{Some General Properties of the Reliability Polynomial}

In this section we review some general properties of the reliability polynomial
that will be used in the rest of the paper.  As is evident from its definition,
the reliability polynomial has the properties that
\beq
R(G,0)=0 \ , \quad R(G,1)=1,
\label{rgp01}
\eeq
\beq
{\rm If} \ \ p \in [0,1], \quad {\rm then} \quad R(G,p) \in [0,1]
\label{rgpbound}
\eeq
and
\beq
\frac{dR(G,p)}{dp} \ge 0 \quad {\rm for} \quad p \in [0,1] \ .
\label{deriv}
\eeq

In general, an edge can be classified as (i) a loop, i.e., an edge that
connects a vertex to itself, (ii) a bridge (= co-loop) with the property that
if it is deleted, this increases the number of connected components of the
graph by one (so that, if $G$ is connected, deleting the bridge breaks $G$ into
two disjoint parts), or (iii) neither a loop nor a bridge.  We shall use the
standard mathematical notation that $G/e$ means the graph obtained by deleting
$e$ and identifying the two vertices that were connected by $e$ (called
contracting $G$ on $e$) and $G-e$ means the graph obtained by deleting $e$.
Now if $e$ is a loop, $e=e_\ell$, then the probability that all terminals are
connected with each other is independent of $e_\ell$, so
\beq
R(G,p)=R(G/e_\ell,p) \ .
\label{reloop}
\eeq
If $e=e_b$ is a
bridge, then the probability that it is present is an overall multiplicative
factor in $R$, so
\beq
R(G,p)=pR(G/e_b,p) \ .
\label{rebridge}
\eeq
If $e$ is neither a loop nor a bridge, then $R(G,p)$ satisfies the
deletion-contraction relation
\beq
R(G,p)=(1-p)R(G-e,p)+pR(G/e,p) \ .
\label{delcon}
\eeq
That is, if the edge $e$ is not present, then $R(G,p)$ has the value given by
the reliability polynomial of a graph $G-e$ times the factor $(1-p)$; if the
edge is present, then $R(G,p)$ has the same value as the reliability polynomial
of a graph contracted on $e$, multiplied by $p$, and since these two
possibilities ($e$ not present or present) are mutually exclusive and exhaust
the possibilities, the full polynomial $R(G,p)$ is the sum of these two
possibilities.

We next comment on some general properties of our $r(\{G\},p)$ function.  We
recall that a tree graph $T_n$ is a connected graph with $n$ vertices and no
circuits.  Since $R(T_n,p)=p^{n-1}$, it follows that $r(\{T\},p)=p$, where
$\{T\} \equiv \lim_{n \to \infty} T_n$.  However, in general, $r$ differs from
$p$ for network topologies that are more complicated than that of a tree graph.
The $r$ function can give insight into the reliability behavior of large
networks.  From obvious arguments, it follows that any reliability function,
whether the $k$-terminal reliability function $R_k(G,p)$ or the all-terminal
reliability $R_A(G,p)$, is a monotonically increasing function of $p$ on the
relevant interval $p \in [0,1]$.  It follows that $r(\{G\},p)$ is also a
monotonically nondecreasing function of $p$ on this interval.

 From the basic definition (\ref{R}) one can generalize $p$ from the physical
interval $0 \le p \le 1$ to arbitrary real or, indeed, complex $p$. This
generalization is necessary when one calculates the zeros of $R(G,p)$ in the
complex $p$ plane.  Using our calculations of $R(G,p)$ for a variety of
families of lattice strip graphs, we shall determine exactly the continuous
accumulation set of the zeros of $R(G,p)$ in the complex $p$ plane as $|V| \to
\infty$ for each family of graphs $G$.  We shall denote this accumulation set
as ${\cal B}$.  In this context we note that in 1992 Brown and Colbourn
conjectured that for an arbitrary connected graph $G$, all of the zeros of the
reliability polynomial $R(G,p)$ lie in the disk $|p-1| \le 1$ \cite{bc} (some
related work is \cite{wagner,sbounds,hpp}).  Recently, Royle and Sokal have
reported counterexamples that show the Brown-Colbourn conjecture to be false
\cite{bcf}, although the maximal values of $|p-1|$ found so far are only
slightly greater than unity.  Extending the counterexamples found in
\cite{bcf}, we have also found several recursive families of graphs for 
which the conjecture is false. 

One can also address a related question for the continuous asymptotic
accumulation set ${\cal B}$: Consider a recursive family of arbitrary connected
graphs $G = G(V,E)$ and, as above, let ${\cal B}$ be the asymptotic
accumulation set of zeros of $R(G,p)$ in the limit $|V| \to \infty$.  Is it
true that all points on ${\cal B}$ are contained in the disk disk $|p-1| \le
1$?  Clearly, although the Brown-Colbourn conjecture is false, the answer to
the above question could still be yes since the zeros of $R(G,p)$ that lie
outside the disk $|p-1|$ might, {\it a priori}, not accumulate to form part of
${\cal B}$.  However, we shall show that the answer to this question is no,
i.e. we shall exhibit recursive families of graphs for which ${\cal B}$ extends
outside the disk $|p-1|$.  Using our exact results, we shall address a number
of other interesting questions concerning ${\cal B}$.  It is also of interest
to note that Brown and Colbourn proved that all of the real roots of the
reliability polynomial lie in the set $\{0\} \cup (1,2]$ \cite{bc}.

\bigskip

Since we shall present calculations for strip graphs with free and periodic or
twisted periodic longitudinal boundary conditions, we observe the following
inequality, where $BC_y$ refers to any given transverse boundary condition

\begin{th} For a strip graph $G_s$ of length $L_x$ and width $L_y$ vertices,
\beq
R(G_s, L_y \times L_x, BC_y,(T)PBCx,p) \ge R(G_s, L_y \times L_x, BC_y,FBC_x,p)
\label{rinequality}
\eeq
\end{th}
\medskip

{\bf Proof} \quad These inequalities follow because for a given type of lattice
strip graph $G_s$ with a given length $L_x$ and width $L_y$, the strip with
periodic or twisted periodic longitudinal boundary conditions has a greater
connectivity than strip with free longitudinal boundary conditions. In
particular, for a given $G_s$, if all of the transverse edges at a fixed
longitudinal location are absent, this does not produce two disjoint graphs for
a cyclic or M\"obius strip but does for the free strip.  $\Box$.

\bigskip

\section{Relations Between the Reliability Polynomial and the Tutte Polynomial
and Potts Model Partition Function}

We recall that the Tutte polynomial of a graph $G=(V,E)$ is
\cite{tutte54}-\cite{boll}
\beq
T(G,x,y)=\sum_{G^\prime \subseteq G} (x-1)^{k(G^\prime)-k(G)}
(y-1)^{c(G^\prime)}
\label{t}
\eeq
where $k(G^\prime)$ and $c(G^\prime)$ denote the number of components and
linearly independent circuits of $G^\prime$, with $c(G^\prime) =
|E^\prime|+k(G^\prime)-|V|$.  Since we deal only with connected graphs here,
$k(G)=1$.  (We follow standard notation for the variables $x$ and $y$ of the
Tutte polynomial and caution the reader that these have no connection with the
$x$ (longitudinal) and $y$ (transverse) axes of the strip graphs.)  The
reliability polynomial $R(G,p)$ can be expressed in terms of the Tutte
polynomial $T(G,x,y)$ as \cite{colbourn}
\beq
R(G,p) = p^{|V|-1}(1-p)^{|E|-|V|+1}T(G,1,\frac{1}{1-p}) \ .
\label{rt}
\eeq
We shall use this relation to calculate reliability polynomials for various
families of graphs from our previous calculations of the Tutte polynomials for
these graphs \cite{a}-\cite{ts}.  We have also used an iterative application
of the deletion-contraction relation (\ref{delcon}) to calculate reliability
polynomials directly.

The Tutte polynomial is equivalent to the partition function of the Potts spin
model.  On a lattice, or more generally, on a graph $G$, at temperature $T$,
this model is defined by the partition function \cite{wurev,wu87}
\beq
Z(G,q,v) = \sum_{ \{ \sigma_i \} } e^{-\beta {\cal H}}
\label{zfun}
\eeq
with the Hamiltonian
\beq
{\cal H} = -J \sum_{\langle i j \rangle} \delta_{\sigma_i \sigma_j}
\label{ham}
\eeq
where $\sigma_i=1,...,q$ are the spin variables on each vertex $i \in G$;
$\beta = (k_BT)^{-1}$; $\langle i j \rangle$ denotes pairs of adjacent
vertices; and we use the notation
\beq
K = \beta J \ , \quad a = e^K \ , \quad v = a-1 \ .
\label{kvdef}
\eeq
This partition function can be written as \cite{fk}
\beq
Z(G,q,v) = \sum_{G^\prime \subseteq G} q^{k(G^\prime)}v^{e(G^\prime)} \ .
\label{cluster}
\eeq
This formula enables one to generalize $q$ from ${\mathbb Z}_+$ to ${\mathbb
R}$ or, indeed, ${\mathbb C}$.  From it one also directly deduces the
equivalence
\beq
Z(G,q,v)=(x-1)^{k(G)}(y-1)^{|V|}T(G,x,y)
\label{zt}
\eeq
where
\beq
x=1+\frac{q}{v}
\label{xqv}
\eeq
and
\beq
y=v+1=e^K
\label{yqv}
\eeq
so that
\beq
q=(x-1)(y-1) \ .
\label{qxy}
\eeq
Combining (\ref{rt}) and (\ref{zt}), we have, for a connected graph $G$, the
relation between the reliability polynomial and the Potts model partition
function
\beq
R(G,p) = (1-p)^{|E|} \ \lim_{q \to 0} \ q^{-1}Z(G,q,v=\frac{p}{1-p}) \ .
\label{rz}
\eeq
Note that $Z(G,q,v)$ always has an overall factor of $q$, which cancels the
factor of $q^{-1}$ in (\ref{rz}).  The valuation $y=1/(1-p)$ in (\ref{rt}), or
equivalently, $v=p/(1-p)$ in (\ref{rz}) can also be expressed as
\beq
p=1-e^{-K} = \frac{v}{1+v}\ .
\label{pek}
\eeq
Thus the physical range of $p \in [0,1]$ for the network corresponds to the
physical range of temperature for the $q=0$ Potts ferromagnet, with $T=0 \
\Leftrightarrow \ p=1$ and $T=\infty \ \Leftrightarrow \ p=0$.  The physical
range of temperature for the Potts antiferromagnet would correspond to the
unphysical interval $-\infty \le p \le 0$ for the network.

\section{Structural Results}

\subsection{General Recursive Families of Graphs}

A general form for the Tutte polynomial for the strip graphs considered here,
or more generally, for recursively defined families of graphs $G_m$ comprised
of $m$ repeated subgraph units, is \cite{a}
\beq T(G_m,q,v) =
\frac{1}{x-1}\sum_{j=1}^{N_{T,G,\lambda}} c_{G,j}(\lambda_{G,j})^m
\label{tgsum}
\eeq
where the terms $\lambda_{G,j}$, the coefficients $c_{G,j}$, and the total
number $N_{Z,G,\lambda}$ depend on $G$ through the type of lattice, its width,
$L_y$, and the boundary conditions, but not on the length.

We observe the following general result:

\begin{th} \quad Consider a recursive family of graphs $G_m$ comprised of $m$
repeated subgraph units.  Then the reliability polynomial $R(G_m,p)$ has the
general form
\beq
R(G_m,p) = \sum_{j=1}^{N_{R,G,\mu}} c_{R,G,j} (\mu_{G,j})^m
\label{rgsum}
\eeq
\end{th}
where the coefficients $c_{R,G,j}$ and terms $\mu_{G,j}$'s depend on
$G$ through the type of lattice, its width, $L_y$, and the boundary conditions,
but not on the length.

\medskip

{\bf Proof} \quad This is proved by combining (\ref{rt}) and (\ref{tgsum}).
Clearly, the $\mu_{G,j}$'s and $c_{R,G,j}$'s can be determined from the
$\lambda_{G,j}$'s and $c_{G,j}$'s.  $\Box$  The reduction of the form
(\ref{tgsum}) to (\ref{rgsum}) is analogous to the reduction that we had 
found for chromatic polynomials of recursive families of graphs 
\cite{w}-\cite{k}. 

\medskip

It is convenient to factor out a power of $p$ and write (\ref{rgsum}) as
\beq
R(G_m,p) = p^{a_{1,G}m + a_{0,G}}
\sum_{j=1}^{N_{R,G,\alpha}} c_{R,G,j} (\alpha_{G,j})^m
\label{rgsumfactored}
\eeq
with
\
\beq
\alpha_{G,j}=p^{-a_{1,G}}\mu_{G,j}
\label{betaalpha}
\eeq
where again the $a_{i,G}$, $i=0,1$, depend on $G$ through the type of lattice,
its width, $L_y$, and the boundary conditions, but not on the length.
Obviously, $N_{R,G,\alpha}=N_{R,G,\mu}$.  We find that for $p=1$, for all of
the families that we have considered, all except one of the $N_{R,G,\alpha}$
$\alpha_{G,j}$'s vanish, and the nonzero term, which can be taken to be the
first, has the value unity:
\
\beq
\alpha_{G,1}=1 \ , \quad \alpha_{G,j}=0 \ , \quad 2 \le j \le N_{R,G,\alpha}
\quad {\rm for} \ \ p=1
\label{alphap1}
\eeq
For the strips of the square and triangular lattices considered here,
\beq
a_{1,G}=L_y  \ , \quad G \ \ {\rm of} \ \ sq, t \ \ {\rm type}
\label{a1gsqtri}
\eeq
and for the strips of the honeycomb lattice,
\beq
a_{1,G}=2L_y  \ , \quad G \ \ {\rm of} \ \ hc \ \ {\rm type}
\label{a1ghc}
\eeq
Note that for all of the strips of the square and triangular lattice with
periodic or twisted periodic longitudinal boundary conditions, $m=L_x$.
For the corresponding strips with free longitudinal boundary conditions, it is
convenient to set $m=L_x-1$;  with this definition, it follows that for strips
with both free and periodic (or twisted periodic) longitudinal boundary
conditions, $m$ is equal to length measured in units of edges, with $L_x$
being the length in terms of vertices.  The situation with the honeycomb strips
is different.  For example, for a cyclic strip of the honeycomb lattice,
envisioned as a brick lattice, $L_x=2m$; that is, if the strip is $m$ bricks
long, then the length, measured in terms of vertices, is $2m$.

Following our earlier nomenclature \cite{w}, we denote a term $\alpha_{G,j}$ in
(\ref{rgsumfactored}) as leading (= dominant) if it has a magnitude greater
than or equal to the magnitude of other $\alpha_{G,j^\prime}$'s.  In the limit
$n \to \infty$ the leading $\alpha_{G,j}$ in $R(G,p)$ determines the function
$r(\{G\},p)$.  The continuous locus ${\cal B}$ where $r(\{G\},p)$ is
nonanalytic thus occurs where there is a switching of dominant $\alpha_{G,j}$'s
in $R$, respectively, and is the solution of the equation of degeneracy
in magnitude
\beq
|\alpha_{G,j}|=|\alpha_{G,j^\prime}|
\label{degeneq}
\eeq
between two dominant $\alpha$'s.  

\subsection{Cyclic Strips of Regular Lattices}

Here we will derive some structural properties of reliability polynomials for
cyclic strips of regular lattices from corresponding properties of Tutte
polynomials, using eq. (\ref{rt}).  In \cite{cf} it was shown that for cyclic
and M\"obius strips of the square lattice of fixed width $L_y$ and arbitrary
length $L_x$ (and also for cyclic strips of the triangular lattice) the
coefficients $c_j$ in the Tutte polynomial are polynomials in $q$ with the
property that for each degree $d$ there is a unique polynomial, denoted
$c^{(d)}$.  Further, this was shown to be
\beq
c^{(d)} = U_{2d}(q^{1/2}/2) = \sum_{j=0}^d (-1)^j {2d-j \choose j}
q^{d-j}
\label{cd}
\eeq
where $U_n(x)$ is the Chebyshev polynomial of the second kind.  A number of
properties of these coefficients were derived in \cite{cf}.  The first few of
these coefficients are
\beq
c^{(0)}=1 \ , \quad c^{(1)}=q-1 \ , \quad c^{(2)}=q^2-3q+1 \ ,
\label{cd012}
\eeq
\beq
c^{(3)}=q^3-5q^2+6q-1 \ .
\label{cd3}
\eeq
Following our earlier work \cite{dg}, we define
\beqs
\kappa^{(d)} & = & c^{(d)} + c^{(d-1)} \cr\cr
& = & \sum_{j=0}^{d-1} (-1)^j { 2d-1-j \choose j} q^{d-j}
\label{kappad}
\eeqs
and
\beq
\bar\kappa^{(d)}=q^{-1}\kappa^{(d)}=
\sum_{j=0}^{d-1} (-1)^j{2d-1-j \choose j} q^{d-1-j} \ .
\label{kappabar}
\eeq
An important property of $\kappa^{(d)}$ is that it always has
the factor $q$; this is evident from (\ref{kappad}).  The first few of these
polynomials are
\beq
\kappa^{(1)}=q \ , \quad \kappa^{(2)}=q(q-2) \ , \quad
\kappa^{(3)}=q(q-1)(q-3) \ .
\label{kapp123}
\eeq
The following special results for $q=0$, i.e., $x=1$, will be useful:
\beq
\lim_{q \to 0} \bar\kappa^{(d)} = (-1)^{d+1}d
\label{kappabarq0}
\eeq
\beq
c^{(d)} = (-1)^d \quad {\rm for} \quad q=0
\label{cdq0}
\eeq
and
\beq
\frac{dc^{(d)}}{dq} = (-1)^{d-1}\frac{(d+1)d}{2} \quad {\rm for} \ \ q=0
\label{diffcdq0}
\eeq
so that, using (\ref{qxy}) to express $c^{(d)}$ as a function of $x$ and $y$,
\beq
\frac{\partial c^{(d)}}{\partial x} = (-1)^{d-1}(y-1)\frac{(d+1)d}{2}
\quad {\rm for} \ \ x=1 \ .
\label{diffcdx1}
\eeq

The terms $\lambda_{T,G_s,L_y,j}$ that occur in (\ref{tgsum}) can be classified
into sets, with the $\lambda_{T,G_s,L_y,j}(x,y)$ in the $d$'th set being
defined as those terms with coefficient $c^{(d)}$.  In Ref. \cite{cf} the
numbers of such terms, denoted $n_T(L_y,d)$, were calculated.  Labelling the
eigenvalues with coefficient $c^{(d)}$ as $\lambda_{T,G_s,L_y,d,j}$ with $1 \le
j \le n_T(L_y,d)$, the Tutte polynomial for a strip graph of length $L_x=m$ of
a regular lattice of type $G_s$ can be written in the form
\beq
T[G_s(L_y \times m; FBC_y,PBC_y),x,y] = \frac{1}{x-1}\sum_{d=0}^{L_y}
c^{(d)} \sum_{j=1}^{n_T(L_y,d)} (\lambda_{T,G_s,L_y,d,j})^m  \ .
\label{tgsumcyc}
\eeq
The property that $T(G,x,y)$ is a polynomial in $x$, which is evident from its
definition (\ref{t}), is not manifest in (\ref{tgsumcyc}) because of the
$1/(x-1)$ prefactor; however, this prefactor is always cancelled in the
evaluation of the sum on the right-hand side of (\ref{tgsumcyc}) for any
particular type of lattice strip graph $G_m$.  Nevertheless, the presence
of this prefactor means that if one calculates the reliability polynomial for a
cyclic strip graph by evaluating the Tutte polynomial for this graph at $x=1$,
one encounters an expression of the form $0/0$ and must use L'Hopital's rule to
evaluate it.  Thus, differentiating numerator and the denominator $1/(x-1)$
from the prefactor and using eqs. (\ref{cdq0}) and (\ref{diffcdx1}), we have
\beqs
& & T[(G_s(L_y \times m; FBC_y,PBC_y),x=1,y] = \sum_{d=1}^{L_y} (-1)^{d-1}(y-1)
\frac{(d+1)d}{2}\sum_{j=1}^{n_T(L_y,d)}(\lambda_{T,G_s,L_y,d,j})^m \cr\cr
& & + m\sum_{d=0}^{L_y} (-1)^d \sum_{j=1}^{n_T(L_y,d)}
(\lambda_{T,G_s,L_y,d,j})^{m-1} \frac{\partial \lambda_{T,G_s,L_y,d,j}}
{\partial x}
\label{tgsumcycxeq1}
\eeqs
where $G_s$ denotes the type of lattice and the right-hand side is evaluated at
$x=1$.  This provides one way of understanding the origin of certain factor of
$m$ in the exact results that we shall give below for reliability polynomials
of strip graphs of length $m$.

The number $n_T(L_y,d)$ of $\lambda$'s with a given coefficient $c^{(d)}$ is
\cite{a}
\beq
n_T(L_y,d)=\frac{(2d+1)}{(L_y+d+1)}{2L_y \choose L_y-d} \quad {\rm for} \ \
0 \le d \le L_y
\label{ntlyd}
\eeq
and zero otherwise.  The total number $N_{T,L_y,\lambda}$ of different terms
$\lambda_{T,L_y,j}$ in eq. (\ref{tgsum}) is \cite{cf}
\beq
N_{T,L_y,\lambda}=\sum_{d=0}^{L_y} n_T(L_y,d) \ .
\label{nttot}
\eeq
For cyclic and M\"obius strips of the square, triangular, and honeycomb
lattices, we calculated this to be \cite{cf,hca}
\beq
N_{T,L_y,\lambda}={2L_y \choose L_y} \ .
\label{nttotcyc}
\eeq

For cyclic and M\"obius strips of regular lattices $G_s$ with arbitrary
$L_y$, eq. (\ref{ntlyd}) shows that there is a unique
$\lambda_{T,G_s,L_y,d}$ corresponding to the coefficient $c^{(d)}$ of
highest degree $d$, namely $d=L_y$.  We have found \cite{a,ta,hca} that this
term is
\beq
\lambda_{T,G_s,L_y,d=L_y} = 1 \ .
\label{lamtutlyly}
\eeq
As indicated, since this eigenvalue is unique, it is not necessary to append a
third index, as with the other $\lambda$'s, and we avoid this for simplicity.
For each of the cyclic strips of regular lattices $G_s$ considered here,
one can derive a general formula for $\alpha_{G_s,L_y,L_y}$ from (\ref{rt})
in conjunction with (\ref{lamtutlyly}), viz.,
\beq
\alpha_{G_s,L_y,L_y}=(1-p)^{\rho_{G_s}}
\label{alphalyly}
\eeq
where
\beq
\rho_{G_s} = \frac{|E|-|V|}{m} \ .
\label{rho}
\eeq
For example, for the cyclic and M\"obius strips of the square (sq) lattice of
width $L_y$ vertices and length $L_x=m$ vertices, $|V|=L_ym$
and $|E|=(2L_y-1)m$, so that $\rho_{sq}=L_y-1$; for the corresponding strips
of the triangular (t) lattice, $|V|$ is the same and $|E|=(3L_y-2)m$, so that
$\rho_t=2(L_y-1)$.  For the cyclic strip of the honeycomb (hc) lattice of width
$L_y$ vertices and length $L_x=2m$ vertices, we have $|V|=2L_ym$ and
$|E|=(3L_y-1)m$ so that $\rho_{hc}=L_y-1$.

We next have

\begin{th} \quad Consider the reliability polynomial for cyclic strips of the
square, triangular, or honeycomb lattices, denoted generically $G_s$, of
width $L_y$ and arbitrary length $L_x$.  The total number of different terms in
(\ref{rgsum}) satisfies the inequality $N_{R,G_s,L_y,\alpha} \le
(1/2)N_{T,G_s,L_y,\lambda}$ and hence
\beq
N_{R,G_s,L_y,\alpha} \le \frac{1}{2}{2L_y \choose L_y} \ .
\label{nrtot}
\eeq
\label{ncyclic}
\end{th}

\medskip

{\bf Proof} \quad The proof makes use of the expression (\ref{rt}) of the
reliability polynomial $R(G_m,p)$ as a special case of the Tutte polynomial
$T(G_m,x,y)$ for $x=1$ and $y=1/(1-p)$, together with our earlier result
(\ref{tgsumcyc}) for the Tutte polynomial of a cyclic strip graph of the
square, triangular, or honeycomb lattice.  Because the prefactor $1/(x-1)$ in
(\ref{tgsumcyc}) is singular at $x=1$ where one evaluates the Tutte polynomial,
it is necessary that
\beq
\left [ \sum_{d=0}^{L_y}c^{(d)} \sum_{j=1}^{n_T(L_y,d)}
(\lambda_{T,G_s,L_y,d,j})^m \right ]_{x=1} = 0
\label{numzero}
\eeq
This vanishing condition holds for arbitrary $m$, which means that it implies
pairwise relations among various terms $\lambda_{T,G_s,L_y,d,j}$.  The only
such relations that can yield an overall factor of $(x-1)$ are relations of the
form $\lambda_{T,G_s,L_y,d,j}=\lambda_{T,G_s,L_y,d-\ell,j^\prime}$ for $1 \le d
\le L_y$ and odd $\ell$ in the range $1 \le \ell \le d$ for a given $d$,
since these yield expressions of the form
\beqs
c^{(d)}(\lambda_{T,G_s,L_y,d,j})^m +
c^{(d-\ell)}(\lambda_{T,G_s,L_y,d-\ell,j^\prime})^m
& = &
(c^{(d)}+c^{(d-\ell)})(\lambda_{T,G_s,L_y,d,j})^m
\label{cdcombination}
\eeqs
Using (\ref{cdq0}), one sees that since $\ell$ is odd, the combination
$c^{(d)}+c^{(d-\ell)}$ vanishes at $q=0$ and hence has a factor of $q$.
Using eq. (\ref{qxy}), one sees that this is sufficient to get an
overall factor of $x-1$ so as to yield the vanishing in eq. (\ref{numzero}) at
$x=1$ or equivalently to cancel the prefactor $1/(x-1)$.  Because of the
pairwise equalities connecting each of the $\lambda$'s when evaluated at $x=1$,
the total number of these terms in the reliability polynomial is reduced to
half of the number for the Tutte polynomial, eq.  (\ref{nttot}). Further
equalities are, in principle, possible and would further reduce the number of
different $\lambda$'s.  This completes the proof of the theorem. $\Box$

\medskip

Two remarks are in order here.  First, we note that in all of our calculations,
we find $\ell=1$; that is, the pairwise equalities among the $\lambda$'s occur
for adjacent values of $d$.  Hence, the linear combination in
eq. (\ref{cdcombination}) is $c^{(d)}+c^{(d-\ell)}=\kappa^{(d)}$, where
$\kappa^{(d)}$ was given in (\ref{kappad}).  Second, in all of our
calculations, we have observed that the inequality (\ref{nrtot}) is realized
as an equality.  Let us illustrate how the pairwise equalities work in some
special cases.  For $L_y=2$, we have $n_T(2,1)=3$ and $n_T(2,0)=2$.  At $x=1$,
the single $\lambda_{T,2,2}$ is equal to one of the three $\lambda_{T,2,1,j}$,
which we denote $\lambda_{T,2,1,3}$.  The other two of these three
$\lambda_{T,2,1,j}$'s are equal to the two respective $\lambda_{T,2,0,j}$'s:
$\lambda_{T,2,1,j}=\lambda_{T,2,0,j}$ for $j=1,2$. For $L_y=3$, we have
$n_T(3,3)=1$, $n_T(3,2)=5$, $n_T(3,1)=9$, and $n_T(3,0)=5$.  At $x=1$,
$\lambda_{T,3,3}$ is equal to one of the five $\lambda_{T,3,2,j}$'s while the
other four are equal to four of the nine $\lambda_{T,3,1,j}$'s, and finally,
the remaining five of these $\lambda_{T,3,1,j}$'s are equal, respectively, to
the five $\lambda_{T,3,0,j}$'s.

\subsection{Cyclic Strips of the Square Lattice with Self-Dual Boundary
Conditions}

We have previously calculated the Tutte polynomial for cyclic strips of the
square lattice with self-dual boundary conditions \cite{jz,dg,sdg}.  These
strip graphs have (i) a fixed transverse width $L_y$, (ii) arbitrarily great
length $L_x$, (iii) periodic longitudinal boundary conditions, and (iv) are
such that each vertex on one side of the strip, which we take to be the upper
side (with the strip oriented so that the longitudinal, $x$ direction is
horizontal) is joined by edges to a single external vertex.  A strip graph of
this type will be denoted generically as $G_D$ (where the subscript $D$ refers
to the self-duality) and, when its size is indicated, as $G_D(L_y \times L_x)$.
For cyclic strips with self-dual boundary conditions we determined the general
structure of the Tutte polynomial in \cite{dg}.  We showed that the
coefficients were precisely the $\kappa^{(d)}$ polynomials. The general form of
the Tutte polynomial is \cite{dg,sdg}
\beq
T(G_D[L_y \times m,cyc.],x,y) = \sum_{d=1}^{L_y+1} \bar\kappa^{(d)}
\sum_{j=1}^{n_T(G_D,L_y,d)} (\lambda_{T,G_D,L_y,d,j})^m
\label{tgsumcycsdg}
\eeq
where $\bar\kappa^{(d)}$ was defined in eq. (\ref{kappabar}) and
\beq
n_T(L_y,d)=\frac{2d}{L_y+d+1}{2L_y+1 \choose L_y-d+1}
\label{ntlydgd}
\eeq
so that for the total number of terms \cite{dg},
\beq
N_{T,G_D,L_y,\lambda} = { 2L_y+1 \choose L_y+1} \ .
\label{nttotgd}
\eeq
Note that, in contrast to the situation with the cyclic strips, because the
Tutte polynomial does not contain any prefactor of $1/(x-1)$, there is no need
to use L'Hopital's rule in evaluating this polynomial to get the reliability
polynomial.  Related to this, the coefficients are $\bar\kappa^{(d)}$, i.e., do
not contain the factor of $q$ that the $\kappa^{(d)}$ coefficients do.  In this
evaluation at $x=1$, i.e., $q=0$, the $\bar\kappa^{(d)}$ simply reduce to the
constants $(-1)^{d+1}d$.  The reliability polynomial for the cyclic self-dual
strip, $R(G_D[L_y \times m],p]$, thus contrasts with the reliability polynomial
for the cyclic strip graphs in not having any terms proportional to $m$, a
consequence of the fact that one did not have to use L'Hopital's rule in
evaluating the corresponding Tutte polynomial at $x=1$.

 Our next result is
\begin{th} The reliability polynomial of the self-dual strip graph
$G_D[L_y \times L_x]$ is (with $L_x = m$)
\beq
R(G_D[L_y \times m],p) = p^{L_ym} \sum_{d=1}^{L_y+1} (-1)^{d+1}d
\sum_{j=1}^{n_T(G_D,L_y,d)}(\alpha_{G_D,L_y,d,j})^m
\label{rgd}
\eeq
{\rm where}
\beq
\alpha_{G_D,L_y,d,j}=(1-p)^{L_y} \left . \lambda_{T,G_D,L_y,d,j} \right |_{
x=1,y=1/(1-p)} \ .
\label{alpha_lambda_gd}
\eeq
\end{th}

\medskip

{\bf Proof} \quad This result follows directly from (\ref{tgsumcycsdg}) and
(\ref{rt}), using (\ref{kappabarq0}) and the fact that the number of vertices
and edges for a graph in this family are $|V|=L_yL_x+1$ and $|E|=2L_yL_x$.
$\Box$

\medskip

 From our explicit calculation of $R(G_D[L_y \times m],p)$ for several values
of the width $L_y$, we find that there is no reduction in the total number of
$\lambda$'s when one carries out the evaluation at $x=1$ for eq. (\ref{rgd}).
Hence, for the cases that we have studied, this total number of $\lambda$'s in
the reliability polynomial for the self-dual cyclic strips of the square
lattice is the same as the number for the Tutte polynomial (Potts model
partition function) for these graphs.  Specifically, this is
$N_{R,G_D,L_y,\lambda}=3, 10, 35$ for $L_y=1,2,3$, respectively.

\subsection{Strips with Free Longitudinal Boundary Conditions}

As before for chromatic and Tutte polynomials, it is convenient to express the
results in terms of a generating function.  For a strip graph of type $G_s$
(where this includes the specification of the transverse boundary conditions)
and length $m$ longitudinal edges, we have
\beq
\Gamma(G_s,p,z) = \sum_{m=0}^\infty R((G_s)_m,p)z^m
\label{gamma}
\eeq
where
\beq
\Gamma(G_s,p,z) = \frac{ {\cal N}(G_s,p,z)}{{\cal D}(G_s,p,z)} \ .
\label{gammazcalc}
\eeq
The numerator and denominator are rational functions of $z$ and $p$ of the form
\beq
{\cal N}(G_s,p,z) = \sum_{j=1}^{{\rm deg}_z({\cal N})} A_{G_s,j} z^j
\label{n}
\eeq
\beq
{\cal D}(G_s,p,z) = 1 + \sum_{j=1}^{{\rm deg}_z({\cal D})} b_{G_s,j} z^j \ .
\label{d}
\eeq
The property (\ref{alphap1}) is manifested in the relations
\beq
b_{G_s,1}=-1 \ , \quad b_{G_s,j}=0 \ , 2 \le j \le N_{G_s,\alpha} \quad
{\rm for} \ \ p=1
\label{bjp1}
\eeq
For the strips of the square and triangular lattices with free longitudinal
boundary conditions,
\beq
A_{G_s,0}=R(T_{L_y},p)= p^{L_y-1} \quad {\rm for} \quad G_s=sq,t
\label{nstrip}
\eeq

\subsection{Accumulation Set of Zeros of $R(G,p)$}

Concerning the continuous accumulation set ${\cal B}$ of the zeros of $R(G,p)$
for cyclic and M\"obius lattice strip graphs in the limit $L_x \to \infty$, we
note

\begin{th} \quad
Consider the lattice strip graphs $G[L_y,L_x,cyc]$ and $G[L_y,L_x,Mb]$.  The
continuous accumulation set ${\cal B}$ defined in the limit $L_x \to \infty$
is the same for both of these.
\label{Bcycmb}
\end{th}

{\bf Proof} \quad This is a corollary of earlier theorems that we have proved
for the accumulation set of the zeros of the Tutte polynomial or equivalent
Potts model partition function, stating that these are the same for a given
strip with cyclic or M\"obius longitudinal boundary conditions.
$\Box$

\medskip

We next discuss the construction of families of graphs $G$ for which the
reliability polynomial $R(G,p)$ has zeros that extend to the circle $|p-1|=1$. 
A useful lemma can be stated as follows \cite{colbourn,bc}: 
Let $G=(V,E)$ be a connected graph with no multiple edges.  We also assume that
$G$ has no loops; this assumption incurs no loss of generality since loops do
not affect the reliability polynomial.  Now let $G_\ell$ be the graph obtained
from $G$ by replacing each edge by $\ell$ edges connecting the same pair of
vertices.  Then
\beq
R(G_\ell,p) = R(G,p^\prime)
\label{rgell}
\eeq
where
\beq
p^\prime = 1-(1-p)^\ell \ .
\label{pprime}
\eeq
In addition to the proofs in \cite{colbourn,bc}, this result can be understood
easily from a physics viewpoint, using the expression of the reliability
polynomial as a special case of the Potts model partition function. Replacing
every edge in $G$ by $\ell$ edges connecting the same vertices has the effect
of replacing the spin-spin coupling $J$ by $J^\prime = \ell J$ in the Potts
model Hamiltonian ${\cal H}$, and hence $K \to K^\prime = \ell K$, so that,
since $y=e^K$ (cf. eq. (\ref{yqv})),
\beq
y \to y^\prime = y^\ell, \ \ i.e., \ \ v \to v^\prime = (v+1)^\ell-1
\label{yell}
\eeq
and
\beq
Z(G_\ell,q,v)=Z(G,q,v^\prime)
\label{zgell}
\eeq
The result (\ref{rgell}) then follows via (\ref{rz}) (or (\ref{rt})). 

As noted in \cite{colbourn,bc}, using this result, one can generate families of
graphs, namely those with replicated edges, that extend to $|p-1|=1$. 
The relation (\ref{zgell}) is equivalent to
\beq
T(G_\ell,x,y) = \left ( \frac{y^\ell - 1}{y-1} \right )^{n-1}
T(G,x^\prime,y^\prime)
\label{tgell}
\eeq
where $y^\prime$ was given in (\ref{yell}) and
\beq
x^\prime = 1 + \frac{(x-1)(y-1)}{y^\ell - 1}
\label{xell}
\eeq
Now let $G=(V,E)$ be a simple graph (i.e. a graph without loops or multiple
edges) that is connected and has at least one edge.  As before, the exclusion
of loops does not incur any loss of generality since loops have no effect on
the reliability polynomial.  Now let $G_\ell$ be the graph obtained from $G$ by
replacing each edge by $\ell$ edges connecting the same pair of vertices.  Then
the zeros of $R(G_\ell,p)$ in the complex $p$ plane include the following:
\beq
p = 1 - e^{\frac{2\pi i k}{\ell}} \ , \quad k=0,1,..., \ell-1
\label{pzero_t2}
\eeq
and these satisfy $|p-1| =1$.  This follows by noting that in general, for the
specified $G$ with at least one edge, $R(G,p)$ contains a factor $p^{|V|-1}$
and hence $G_\ell$ contains a factor of $(p^\prime)^{|V|-1}$.  Therefore
$R(G_\ell,p)$ has at least one zero at $p^\prime =0$.  Recalling the relation
(\ref{pprime}) and solving for the $\ell$ roots of this equation in the $p$
plane, we obtain (\ref{pzero_t2}), which satifies $|p-1|=1$.

Let us give an example.  Perhaps the simplest is to consider the graph
$G=T_2$. Replacing the single edge in this graph by $\ell$ edges yields a graph
commonly denoted the ``thick link'' or ``fat link'', $TL_\ell$.  This is the
(planar) dual to the circuit graph $C_\ell$; i.e., $TL_\ell = C_\ell^*$. From
the elementary result $R(T_2,p)=p$ and our theorem (\ref{rgell}), one gets
$R(FL_\ell,p)=p^\prime$, which has zeros at the points (\ref{pzero_t2}).  The
location of these zeros for $\ell=(a) 2, (b) 3, and (c) 4$ are (a) $p=0,2$; (b)
$p=0$ and $p=3/2 \pm (i/2)\sqrt{3}$; (c) $p=0,2$, and $1 \pm i$. In the limit
$\ell \to \infty$, one then obtains an accumulation set ${\cal B}$ that is the
circle $|p-1|=1$.  Note that in this example, although the number of edges
goes to infinity, the number of vertices is fixed at 2.  One can also consider
families of graphs with a subset of the vertices replicated.  The family of
$L_y=2$ strips of the square lattice with cylindrical or torus boundary
conditions are of this type, with double vertical and single horizontal edges.
These families will be analyzed below.

\section{Calculations for Specific Families of Graphs}

\subsection{$L_y=1$ Strips of the Square Lattice}

The reliability function for the free $L_y=1$ strip is a special case of the
result for a tree graph $T_m$ with $m$ vertices, namely $R=p^{m-1}$.
For the case of cyclic boundary conditions, the strip is the circuit graph,
$C_m$.  By an elementary application of the deletion-contraction theorem, one
has
\beq
R(C_m,p) = p^{m-1}[m(1-p)+p] \ .
\label{Rcn}
\eeq
Hence,
\beq
r=p
\label{rsqxy1}
\eeq
for the infinite-length limits of both the free and cyclic $L_y=1$ strips.
The polynomial $R(C_m,p)$ has a zero of multiplicity $m-1$ at $p=0$ and another
zero at $p=m/(m-1)$, which monotonically approaches $p=1$ from above as $m \to
\infty$.  The $r$ function (\ref{rsqxy1}) for these $L_y=1$ strips has the
special property that $dr/dp=1$.  As will be seen, the $r$ functions that we
shall calculate for other strips have various nonzero slopes at $p=0$ but, in
contrast to (\ref{rsqxy1}), they have the derivative $dr/dp=0$ at $p=1$.

\bigskip

\subsection{$L_y=2$ Strip of the Square Lattice with Free Boundary Conditions}

The generating function for this strip is of the form (\ref{gammazcalc}) with
${\cal N}=p$ and
\beqs
{\cal D}(p,z) & = & 1-p^2(4-3p)z+p^4(1-p)z^2 \cr\cr
              & = & (1-p^2 \alpha_{sq,2,0,1}z)(1-p^2 \alpha_{sq,2,0,2}z)
\label{dengamma}
\eeqs
where
\beq
\alpha_{sq,2,0,j} = \frac{1}{2}\left [ 4-3p \pm \sqrt{R_2} \ \right ]
\ , \quad j=1,2
\label{alpha20j}
\eeq
with
\beq
R_2=12-20p+9p^2 \ .
\label{r2}
\eeq
The subscripts in the notation make reference to the structure of the
$\alpha$'s for the cyclic strip; see below.  We observe that $R(sq[2 \times
m],p)$ has the factor $(4-3p)$ if $m \ge 1$ is odd.  For the cases that we have
examined, besides the real root at $p=0$, this reliability polynomial has a
single real root at $p=4/3$ if and only if $m \ge 1$ is odd.

The dominant term in the physical interval $p \in [0,1]$ is
$\alpha_{sq,2,0,1}$, and hence the asymptotic reliability per vertex for $L_x
\to \infty$ is
\beqs
r(sq[2 \times \infty, free],p) & = & p(\alpha_{sq,2,0,1})^{1/2} \cr\cr
& = & \frac{p}{\sqrt{2}} \left [ 4-3p + \sqrt{12-20p+9p^2} \ \right ]^{1/2} \ .
\label{rsqxy2}
\eeqs
This has the derivatives
\beq
\left . \frac{dr}{dp} \right |_{p=0} = \frac{1+\sqrt{3}}{\sqrt{2}} = 1.93185...
\label{drdp_sqxy2_p0}
\eeq
and
\beq
\left . \frac{dr}{dp} \right |_{p=1} = 0 \ .
\label{drdp1}
\eeq

\begin{figure}[hbtp]
\centering
\leavevmode
\epsfxsize=4.0in
\begin{center}
\leavevmode
\epsffile{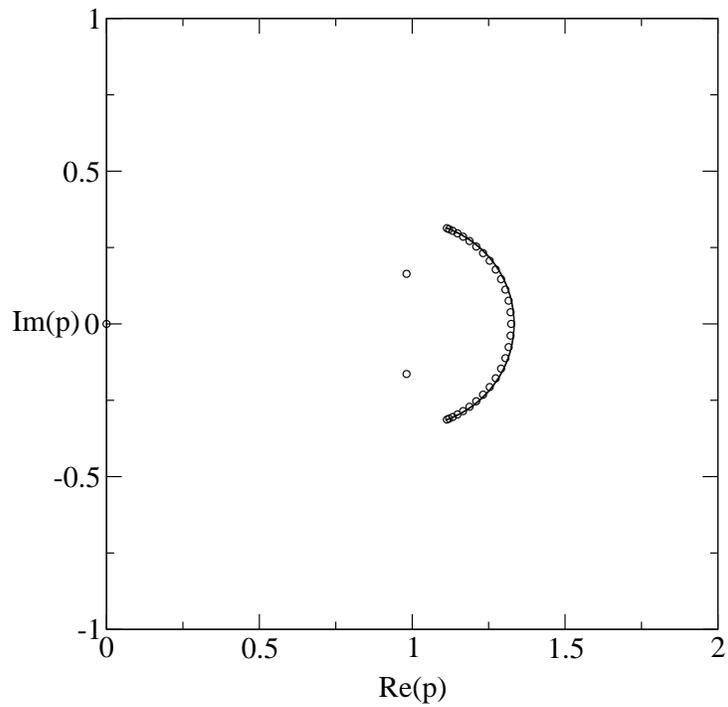}
\end{center}
\caption{\footnotesize{Singular locus ${\cal B}$ for the $L_x \to \infty$ limit
of $sq(2 \times L_x)$ for either free, periodic, or twisted periodic
longitudinal boundary conditions.  For comparison, zeros of the reliability
polynomial are shown for the cyclic strip with $L_x=30$ (i.e., $n=60$).}}
\label{sqpxy2}
\end{figure}

The locus ${\cal B}$ for the $L_y=2$ free strips of the square lattice, shown
in Fig. \ref{sqpxy2}, is formed as the continuous accumulation set of zeros of
the reliability polynomial in the limit of infinite strip length, $m \to
\infty$ and is the solution of the equation expressing the degeneracy of
magnitudes $|\alpha_{sq,2,0,1}|=|\alpha_{sq,2,0,2}|$.  This is an arc of a
circle with radius 1/3 centered at 1,
\beq
p = 1 + \frac{1}{3}e^{i\theta} \quad \theta
\in [-\theta_{sq2}, \theta_{sq2}]
\label{pcircsq2}
\eeq
where
\beq
\theta_{sq2} = {\rm arctan}(2\sqrt{2} \ ) \simeq 70.53^\circ \ .
\label{theta_sq2}
\eeq
This circular arc crosses the real $p$ axis at a single point, which is thus
$p_c$, namely
\beq
p_c = \frac{4}{3} \quad {\rm for} \quad sq, \ FBC_y, \ 2 \times \infty \ .
\label{pcross_sq2}
\eeq
This applies here for the case of free longitudinal boundary conditions, and we
shall show below that it also applies for the case of periodic or twisted
periodic longitudinal boundary conditions.  We find that this property that the
value of $p_c$ for a strip of a given type of lattice and a given set of
transverse boundary conditions is independent of the longitudinal boundary
conditions also holds for other strips, and we indicate this when listing
these values of $p_c$ (or $(p_c)_{eff}$) by displaying only the transverse
boundary condition ($FBC_y$ in eq. (\ref{pcross_sq2})).  The arc has endpoints
at the values of $p$ where the the expression $12-20p+9p^2$ in the square roots
in $\alpha_{sq,2,0,j}$ vanishes so that $\alpha_{sq,2,0,1}=\alpha_{sq,2,0,2}$.
These endpoints are
\beq
p = 1 + \frac{1}{3}e^{\pm i\theta_{sq2}} =
\frac{2}{9}(5 \pm \sqrt{2} \ i) \simeq 1.11111 \pm 0.31427i \ .
\label{pendsqly2}
\eeq
Recall that two roots of an algebraic equation are equal where the discriminant
vanishes; for the equation $\alpha^2-(4-3p)\alpha+1-p=0$ that yields
$\alpha_{sq,2,0,j}$, $j=1,2$ as its roots, this discriminant is $12-20p+9p^2$.

\subsection{$L_y=3$ Strip of the Square Lattice with Free Boundary Conditions}

The generating function for this strip follows from our calculation of
$Z(G,q,v)$ for this strip \cite{s3a} and has the form (\ref{gammazcalc}) with
\beq
{\cal N}=p^2\biggl [ 1 + p^4(1-p)z - p^6(1-p)^3z^2 \biggr ]
\label{numsqy3}
\eeq
and
\beqs
{\cal D}(p,z) & = & 1+\sum_{j=1}^4 b_{sq3f,j}z^j \cr\cr
              & = & \prod_{j=1}^4(1-p^3\alpha_{sq3f,j}z)
\label{densqy3}
\eeqs
where
\beq
b_{sq3f,1}= -p^3(15-24p+10p^2)
\label{bsq3f1}
\eeq
\beq
b_{sq3f,2}=p^6(1-p)(32-66p+46p^2-11p^3)
\label{bsq3f2}
\eeq
\beq
b_{sq3f,3}=-p^9(1-p)^3(15-20p+7p^2)
\label{bsq3f3}
\eeq
\beq
b_{sq3f,4}=p^{12}(1-p)^5
\label{bsq3f4}
\eeq

\begin{figure}[hbtp]
\centering
\leavevmode
\epsfxsize=4.0in
\begin{center}
\leavevmode
\epsffile{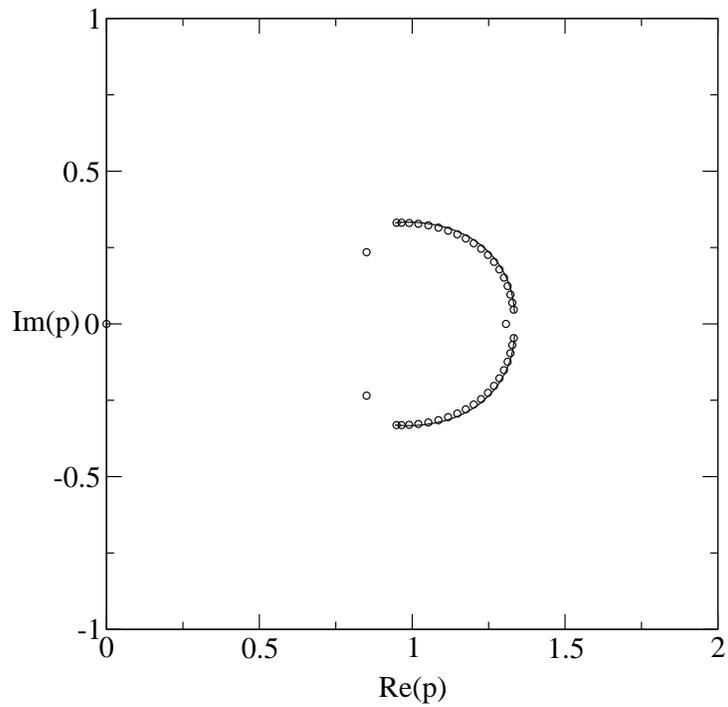}
\end{center}
\caption{\footnotesize{Singular locus ${\cal B}$ for the $L_x \to \infty$ limit
of $sq(3 \times L_x)$ for either free, periodic, or twisted periodic
longitudinal boundary conditions.  For comparison, zeros of the reliability
polynomial are shown for the cyclic strip of this width with $L_x=20$ (i.e.,
$n=60$).}}
\label{sqpxy3}
\end{figure}

In Fig. \ref{sqpxy3} we show the locus ${\cal B}$ for the $L_y=3$ strips
of the square lattice with any longitudinal boundary conditions, together with
zeros for the cyclic strip of this width with $L_x=20$.  This locus consists of
two complex-conjugate arcs which do not cross the real axis.  From the arc
endpoints closest to the real axis, one can infer the effective quantity
$(p_c)_{eff} \simeq 1.335$.   The locus is again concave to the left and
roughly centered about the point $p=1$.

\subsection{$L_y=2$ Cyclic and M\"obius Strips of the Square Lattice}

For the reader's convenience, we shall give some details of our calculation for
this family of graphs.  The Tutte polynomials for these families of graphs were
computed in \cite{a} and have $N_{T,2,\lambda}=6$, in accordance with the
general formula (\ref{nttotcyc}).  There are $n_T(2,0)=2$ terms with
coefficient $c^{(0)}$, denoted $\lambda_{T,sq,2,0,j}$, $j=1,2$;
$n_T(2,1)=3$ terms
with coefficient $c^{(1)}$, denoted $\lambda_{T,sq,2,1,j}$, and the unique
term
with coefficient $c^{(2)}$, which term is $\lambda_{T,sq,2,2}=1$ as a special
case of the general result (\ref{lamtutlyly}).  Explicitly,
\beqs
& & T(sq[L_y=2,L_x=m,cyc],x,y) = \frac{1}{x-1}\sum_{d=0}^2
c^{(d)} \sum_{j=1}^{n_T(2,d)} (\lambda_{T,sq,2,d,j})^m \cr\cr
& & = \frac{1}{x-1}\left [ \sum_{j=1}^2 (\lambda_{T,sq,2,0,j})^m
+ c^{(1)}\sum_{j=1}^3 (\lambda_{T,sq,2,1,j})^m
+ c^{(2)}(\lambda_{T,sq,2,2})^m \right ]
\label{tlad}
\eeqs
where $n_T(L_y,d)$ was recalled from \cite{cf} in eq. (\ref{ntlyd}) and
\cite{a}
\beq
\lambda_{T,sq,2,0,j} = \frac{1}{2}\biggl [ (1+y+x+x^2) \pm \Bigl (
 y^2 +2y(1+x-x^2) + (x^2+x+1)^2 \Bigr )^{1/2} \ \biggr ]
\label{lam20j}
\eeq
\beq
\lambda_{T,sq,2,1,j} = \frac{1}{2}\biggl [x+y+2 \pm \Bigl ( (x-y)^2 +
4(x+y+1) \Bigr )^{1/2} \ \biggr ]
\label{lam21j}
\eeq
with $j=1,2$ corresponding to $\pm$, and
\beq
\lambda_{T,sq,2,1,3}=x \ .
\label{lam213}
\eeq
We have discussed above how equalities occur between certain
$\lambda_{T,G,L_y,d,j}$'s when evaluated at $x=1$.  We show this explicitly
here. First, we have
\beq
\lambda_{T,sq,2,0,j} = \lambda_{T,sq,2,1,j} = \frac{1}{2} \left [y+3 \pm
\sqrt{y^2+2y+9} \ \right ] \quad {\rm for} \ \ x=1
\label{lamlady}
\eeq
where $j=1,2$ refer correspond to $\pm$ on the right-hand side,
respectively. Furthermore,
\beq
\lambda_{T,sq,2,1,3} = \lambda_{T,sq,2,2} = 1 \quad {\rm for} \ \ x=1 \ .
\label{lamladlast}
\eeq
Hence, $N_{R,sq,2,\lambda}=(1/2)N_{T,sq,2,\lambda} = 3$, in accordance with the
inequality (\ref{nrtot}) (realized as an equality) in our theorem above.  Using
L'Hopital's rule to evaluate the Tutte polynomial (\ref{tlad}) at $x=1$, we
obtain
\beqs
& & R(sq[L_y=2,m,cyc],p)=p^{2m} \Biggl [
(\alpha_{sq,2,0,1})^m+(\alpha_{sq,2,0,2})^m - 2(\alpha_{sq,2,2})^m \cr\cr & & +
m p^{-1}(1-p)^2 \Biggl \{ (\alpha_{sq,2,0,1})^{m-1} \left (1 +
\frac{3(1-p)}{\sqrt{R_2}} \right ) \cr\cr
& & + (\alpha_{sq,2,0,2})^{m-1} \left (1 -
\frac{3(1-p)}{\sqrt{R_2}} \right ) - (\alpha_{sq,2,2})^{m-1} \Biggr \} \Biggr ]
\label{Rsqpxy2}
\eeqs
where $R_2$ was defined above in eq. (\ref{r2}), the $\alpha_{sq,2,0,j}$ were
given in (\ref{alpha20j}) above, and
\beq
\alpha_{sq,2,2}=1-p
\label{alpha22}
\eeq
as a special case of (\ref{alphalyly}) above.

For the $L_y=2$ M\"obius (Mb) strip of the square lattice \cite{a}
\beqs
& & T(sq[L_y=2,L_x=m,Mb],x,y) =
\frac{1}{x-1}\Biggl [ \sum_{j=1}^2 (\lambda_{T,sq,2,0,j})^m \cr\cr
& & + c^{(1)}\biggl ( -(\lambda_{T,sq,2,1,1})^m + \sum_{j=2}^3
(\lambda_{T,sq,2,1,j})^m \biggr ) - (\lambda_{T,sq,2,2})^m \Biggr ] \ .
\label{tmb}
\eeqs
 From this we obtain
\beqs
& & R(sq[L_y=2,m,Mb],p)=p^{2m} \Biggl [ (\alpha_{sq,2,0,1})^m +
(\alpha_{sq,2,0,2})^m - (\alpha_{sq,2,2})^m \cr\cr
& & + m p^{-1}(1-p)^2 \Biggl \{
(\alpha_{sq,2,0,1})^{m-1} \left (1 + \frac{3(1-p)}{\sqrt{R_2}} \right ) +
(\alpha_{sq,2,0,2})^{m-1} \left (1 - \frac{3(1-p)}{\sqrt{R_2}} \right )
\cr\cr
& & + (\alpha_{sq,2,2})^{m-1} \Biggr \} \Biggr ]  \ .
\label{Rsqpxy2mb}
\eeqs

For both of these $L_y=2$ cyclic and M\"obius strips of the square lattice, the
dominant term in the physical interval $p \in [0,1]$ is $\alpha_{sq,2,0,1}$ and
hence the asymptotic reliability per vertex, $r$, is the same as that for the
corresponding strip with free longitudinal boundary conditions, given by
(\ref{rsqxy2}):
\beqs
r(sq[2 \times \infty, cyc.],p) & = & r(sq[2 \times \infty, Mb.],p) \cr\cr
                               & = & r(sq[2 \times \infty, free],p) \cr\cr
                               & = & r(sq[2 \times \infty, FBC_y],p) \ .
\label{r2sq}
\eeqs
For the infinite-length limit of these strip graphs, and for the
respective infinite-length limits of other strip graphs to be considered below,
we find that the $r$ function is independent of the longitudinal boundary
conditions.  This is indicated in the last line of eq. (\ref{r2sq}) and
analogous equations below by listing only the type of transverse boundary
condition.  The dominant term for large positive $p$ is $\alpha_{sq,2,0,2}$.
Because the two dominant $\alpha$'s in the reliability polynomials for these
cyclic and M\"obius strips are the same as the two $\alpha$'s that enter in the
reliability polynomial for the corresponding strip with free boundary
conditions, it follows that the locus ${\cal B}$ is identical for the $L_y=2$
strips of the square lattice with free, cyclic, and M\"obius boundary
conditions.  This locus is shown in Fig. \ref{sqpxy2}.

We find that the reliability polynomials $R(sq[2 \times L_x, cyc.],p)$ and
$R(sq[2 \times L_x, Mb],p)$ have only $p=0$ as a real root if $L_x \ge 1$ is
odd, while for even $L_x \ge 2$, they each have an additional real root.  In
Table \ref{sq2realroots} we list the values of this additional respective real
root for the reliability polynomials of these two strips.  The results in this
table suggest that as $L_x \to \infty$, these two respective additional real
zeros for the cyclic and M\"obius strips could approach the same limit.  The
value of this limit could be about 1.3.

\begin{table}
\caption{\footnotesize{Values of the real root (aside from $p=0$) of
$R(sq[2 \times L_x,cyc.],p)$ and $R(sq[2 \times L_x,Mb],p)$ for even $L_x$.}}
\begin{center}
\begin{tabular}{|c|c|c|}
\hline\hline $L_x$ & cyc.  & Mb. \\ \hline\hline
2   & 2.000000  & 1.626538 \\ \hline
4   & 1.404813  & 1.388652 \\ \hline
6   & 1.349507  & 1.346559 \\ \hline
8   & 1.333333  & 1.332635 \\ \hline
10  & 1.327103  & 1.326919 \\ \hline
12  & 1.324413  & 1.324362 \\ \hline
14  & 1.323229  & 1.323214 \\ \hline
16  & 1.322761  & 1.322757 \\ \hline
18  & 1.322658  & 1.322657 \\ \hline
20  & 1.322749  & 1.322748 \\ \hline\hline
\end{tabular}
\end{center}
\label{sq2realroots}
\end{table}

\subsection{$L_y=2$ Cylindrical Strips of the Square Lattice}
\label{sqcyl2}

\begin{figure}[hbtp]
\centering
\leavevmode
\epsfxsize=4.0in
\begin{center}
\leavevmode
\epsffile{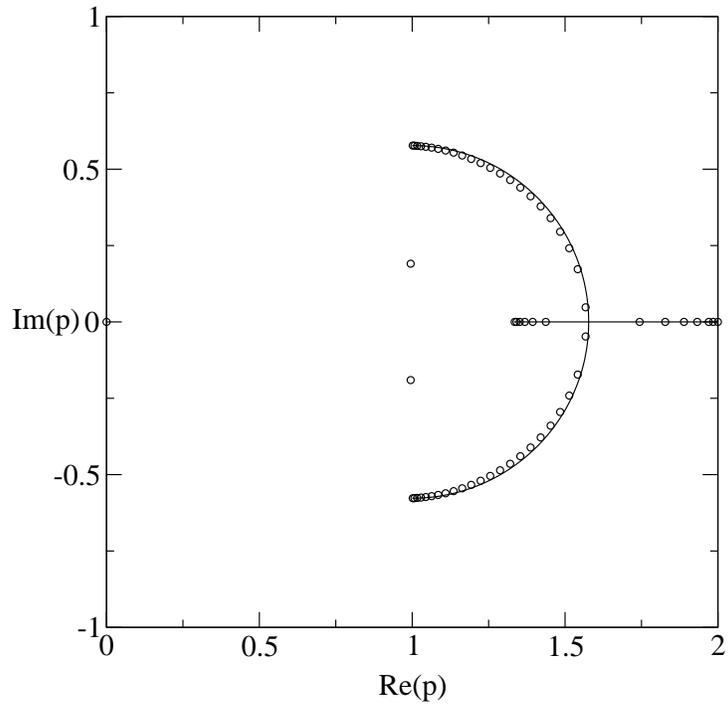}
\end{center}
\caption{\footnotesize{Singular locus ${\cal B}$ for the $L_x \to \infty$ limit
of $sqcyl(2 \times L_x)$ for either free or (twisted) periodic longitudinal
boundary conditions.  For comparison, zeros of the reliability polynomial are
shown for the torus strip with $L_x=30$ (i.e., $n=60$).}}
\label{sqpxpy2}
\end{figure}

This is the minimal-width strip of the square lattice with cylindrical boundary
conditions and involves double vertical edges.  The results for the
reliability polynomial are conveniently expressed in terms of the generating
function (\ref{gamma}) and (\ref{gammazcalc}).  We find
\beq
{\cal N} = R(C_2,p) = p(2-p)
\label{numsqpy2}
\eeq
\beqs
{\cal D} & = & 1-p^2(6-8p+3p^2)z+(1-p)^2p^4z^2 \cr\cr
         & = & \prod_{j=1}^2(1-p^2 \alpha_{sqcyl2,j}z)
\label{densqpy2}
\eeqs
where
\beq
\alpha_{sqcyl2,j} = \frac{1}{2}\left [ 6-8p+3p^2 \pm
\sqrt{(2-p)(4-3p)(4-6p+3p^2)} \ \right ] \ .
\label{alpha_sqcyl2}
\eeq
We observe that $R(sq[2 \times L_x,cyl.],p)$ always has the factor $(2-p)$;
this is associated with the fact that this strip has double vertical edges.
For odd $m \ge 1$, i.e., even $L_x \ge 2$, $R(sq[2 \times L_x,cyl.],p)$ also
has the factor $(3p^2-8p+6)$, with roots $(1/3)(4 \pm \sqrt{2}i)$.  For $m \ge
4$, we observe that $R(sq[2 \times L_x,cyl.],p)$ has, besides the real zeros at
$p=0,2$, a set of real zeros in the interval $(1,2]$.  Among other features, we
note that as $m$ increases, (i) the number of real zeros in this interval
increases; (ii) the smallest zero in this interval decreases monotonically
toward 4/3; (iii) the largest zero increases monotonically toward 2 as $m$
increases, and (iv) the density of zeros is largest in the regions slightly
above 4/3 and slightly below 2.  A plot of these zeros is shown in
Fig. \ref{sqpxpy2}; as is evident from this plot, as $m \to \infty$, these real
zeros away from $p=0$ accumulate to form the line segment on ${\cal B}$
extending between $p=4/3$ and $p=2$.

In the physical interval $p \in [0,1]$ the dominant term is
$\alpha_{sqcyl2,1}$, so the asymptotic reliability per vertex is
\beqs
& & r(sq[2 \times \infty,PBC_y],p) = p(\alpha_{sqcyl2,1})^{1/2} \cr\cr
& = & \frac{p}{\sqrt{2}} \left [ 6-8p+3p^2 + \sqrt{(2-p)(4-3p)(4-6p+3p^2)} \
\right ]^{1/2}
\label{rsqpy2}
\eeqs
This has derivatives
\beq
\left . \frac{dr}{dp} \right |_{p=0} = 1+\sqrt{2} = 2.41421...
\label{drdp_sqpy2_p0}
\eeq
and $dr/dp=0$ at $p=1$.

The locus ${\cal B}$, shown in Fig. \ref{sqpxpy2}, is given by the union of
an arc of a circle with a line segment on the real axis:
\beq
{\cal B}: \quad \{ p = 1 + \frac{1}{\sqrt{3}}e^{i\theta} \ , \quad \theta \in
[-\pi/2, \pi/2] \} \ \cup \{ \frac{4}{3} \le p \le 2 \} \ .
\label{b_sqcpxy2}
\eeq
Hence,
\beq
p_c=2 \quad {\rm for} \quad sq, \ PBC_y, \ 2 \times \infty
\label{pc_sqpy2}
\eeq
The circular arc intersects the real axis and line segment at
$p=1+\frac{1}{\sqrt{3}} \simeq 1.577$.  The endpoints of the arc are located at
\beq
p_{sqce}, p_{sqce}^* = 1 \pm \frac{i}{\sqrt{3}}
\label{psqce}
\eeq
These points, together with the endpoints of the line segment, $p=4/3$ and
$p=2$ are the zeros of the expression in the square root in
$\alpha_{sqcyl2,j}$ in eq. (\ref{alpha_sqcyl2}), which expression is the
discriminant of the equation $\alpha^2-(6-8p+3p^2)\alpha+(1-p)^2=0$ whose roots
are these $\alpha$'s.

\subsection{$L_y=2$ Torus and Klein Bottle Strips of the Square Lattice}

Either using a direct calculation or as a special case of our previous
calculation of the Tutte polynomial \cite{hca}, using (\ref{rt}),
we find
\beqs & & R(sq[L_y=2,m,torus],p)=p^{2m} \Biggl [
(\alpha_{sqtor2,1})^m+(\alpha_{sqtor2,2})^m - 2(\alpha_{sqtor2,3})^m
\cr\cr & & -
m p^{-1}(1-p)^2 \left \{ \frac{(\alpha_{sqtor2,1})^{m-1}}{2} \left (2p-3 +
\frac{(p-2)(6p^2-13p+8)}{\sqrt{(p-2)(3p-4)(3p^2-6p+4)}} \right )
\right. \cr\cr & &
\left. + \frac{(\alpha_{sqtor2,2})^{m-1}}{2} \left (2p-3 -
\frac{(p-2)(6p^2-13p+8)}{\sqrt{(p-2)(3p-4)(3p^2-6p+4}} \right )
+ (1-p)(\alpha_{sqtor2,3})^{m-1} \right \} \Biggr ] \cr\cr & &
\label{Rsqpxpy2}
\eeqs
where
\beq
\alpha_{sqtor2,j}=\alpha_{sqcyl2,j} \quad j=1,2
\label{sqcyltor}
\eeq
with $\alpha_{sqcyl2,j}$, $j=1,2$, given above in (\ref{alpha_sqcyl2}) and,
\beq
\alpha_{sqtor2,3}=(1-p)^2 \ .
\label{alpha_sqc_22}
\eeq

For the $L_y=2$ Klein bottle (Kb.) strip of the square lattice
\beqs
& & R(sq[L_y=2,m,Kb.],p)=p^{2m} \Biggl [
(\alpha_{sqtor2,1})^m+(\alpha_{sqtor2,2})^m - (\alpha_{sqtor2,3})^m
\cr\cr & & -
m p^{-1}(1-p)^2 \left \{ \frac{(\alpha_{sqtor2,1})^{m-1}}{2} \left (2p-3 +
\frac{(p-2)(6p^2-13p+8)}{\sqrt{(p-2)(3p-4)(3p^2-6p+4)}} \right )
\right. \cr\cr & &
\left. + \frac{(\alpha_{sqtor2,2})^{m-1}}{2} \left (2p-3 -
\frac{(p-2)(6p^2-13p+8)}{\sqrt{(p-2)(3p-4)(3p^2-6p+4)}} \right ) -
(1-p)(\alpha_{sqtor2,3})^{m-1} \right \} \Biggr ] \cr\cr & &
\label{Rsqpxpy2kb}
\eeqs

In the physical region $p \in [0,1]$, the dominant term is $\alpha_{sqtor2,1}$
for both of these families of strips, and hence $r(sq[2 \times
\infty,torus/Kb.],p) = r(sq[2 \times \infty,cyl.],p)$.  This equality has
already been incorporated in our notation $r(sq[2 \times \infty,PBC_y],p)$ in
eq. (\ref{rsqpy2}).  This is another example of the feature that for a given
type of strip graph $G_s$ with a given width and set of transverse boundary
conditions, $r$ is independent of the longitudinal boundary conditions.

Similarly, since the locus ${\cal B}$ is determined by the equality in
magnitude of the two dominant eigenvalues $\alpha_{sqtor2,1}$ and
$\alpha_{sqtor2,2}$, and since these are the same for the corresponding strip
with cylindrical boundary conditions, it follows that this locus is the same as
the locus for the $L_y=2$ cylindrical strip of the square lattice, shown in
Fig. \ref{sqpxpy2}.

\subsection{$L_y=3$ Cylindrical Strips of the Square Lattice}
\label{sqcyl3}

Again we express the results in  terms of the generating function for the
reliability polynomial, (\ref{gamma}) and (\ref{gammazcalc}).  We find
\beq
{\cal N} = R(C_3,p) + p^5(3-p)(1-p)^2z = p^2[3-2p+p^3(3-p)(1-p)^2z]
\label{numsqpy3}
\eeq
where $R(C_3,p)=p^2(3-2p)$, and
\beqs
{\cal D} & = &
1-p^3(24-56p+46p^2-13p^3)z+p^6(1-p)^2(24-46p+30p^2-7p^3)z^2-p^9(1-p)^5z^3
\cr\cr
         & = & \prod_{j=1}^3(1-p^3 \alpha_{sqcyl3,j}z) \ .
\label{densqcyl}
\eeqs
The largest of the roots of the associated equation
\beqs
& & \xi^3-(24-56p+46p^2-13p^3)\xi^2 \cr\cr
& & +(1-p)^2(24-46p+30p^2-7p^3)\xi-(1-p)^5=0
\label{xieqsqcyl3}
\eeqs
is the dominant $\alpha_{sqcyl3,max}$, so that
\beq
r(sq[3 \times \infty, PBC_y],p)=p(\alpha_{sqcyl3,max})^{1/3} \ .
\label{rsqcyl3}
\eeq
This function has the property that $dr/dp=2.84207...$ at $p=0$ and $dr/dp=0$
at $p=1$.  We find that the reliability polynomial $R(sq[3 \times \infty,
cyl.],p)$ has only $p=0$ as a real root if $m \ge 1$ is odd, i.e. $L_x \ge 2$
is even, while for even $m \ge 0$, i.e., odd $L_x \ge 1$, it has an additional
real root which decreases as $m$ increases and appears to approach a limit of
roughly 1.4.  In Table \ref{sqcyl3realroots} we list the values of this
additional real root.

\begin{table}
\caption{\footnotesize{Values of the real root (aside from $p=0$) of
$R(sq[3 \times \infty, cyl.],p)$ for odd $L_x$.}}
\begin{center}
\begin{tabular}{|c|c|}
\hline\hline $L_x$ & root \\ \hline\hline
1   & 1.500000   \\ \hline
3   & 1.440337   \\ \hline
5   & 1.422909   \\ \hline
7   & 1.416297   \\ \hline
9   & 1.412887   \\ \hline
11  & 1.410809   \\ \hline
13  & 1.409409   \\ \hline
15  & 1.408403   \\ \hline
17  & 1.407644   \\ \hline
19  & 1.407051   \\ \hline
21  & 1.406575   \\ \hline\hline
\end{tabular}
\end{center}
\label{sqcyl3realroots}
\end{table}

\begin{figure}[hbtp]
\centering
\leavevmode
\epsfxsize=4.0in
\begin{center}
\leavevmode
\epsffile{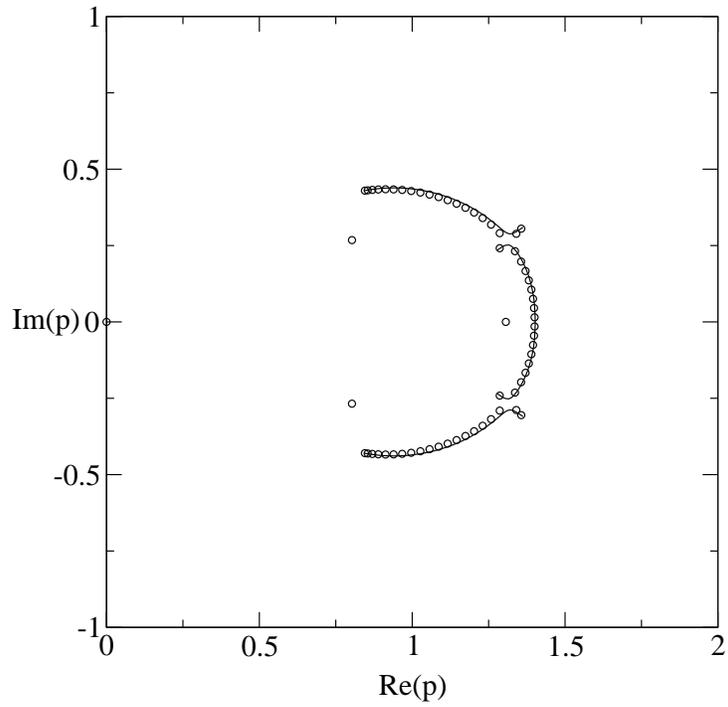}
\end{center}
\caption{\footnotesize{Singular locus ${\cal B}$ for the $L_x \to \infty$ limit
of $sqcyl(3 \times L_x)$ for either free or (twisted) periodic longitudinal
boundary conditions.  For comparison, zeros of the reliability polynomial are
shown for the toroidal strip with $L_x=20$ (i.e., $n=60$).}}
\label{sqpxpy3}
\end{figure}

In Fig. \ref{sqpxpy3} we show a plot of ${\cal B}$ for the $L_x \to \infty$
limit of the $L_y=3$ strip of the square lattice with cylindrical boundary
conditions.

\subsection{$L_y=4$ Cylindrical Strip of the Square Lattice}

\begin{figure}[hbtp]
\centering
\leavevmode
\epsfxsize=4.0in
\begin{center}
\leavevmode
\epsffile{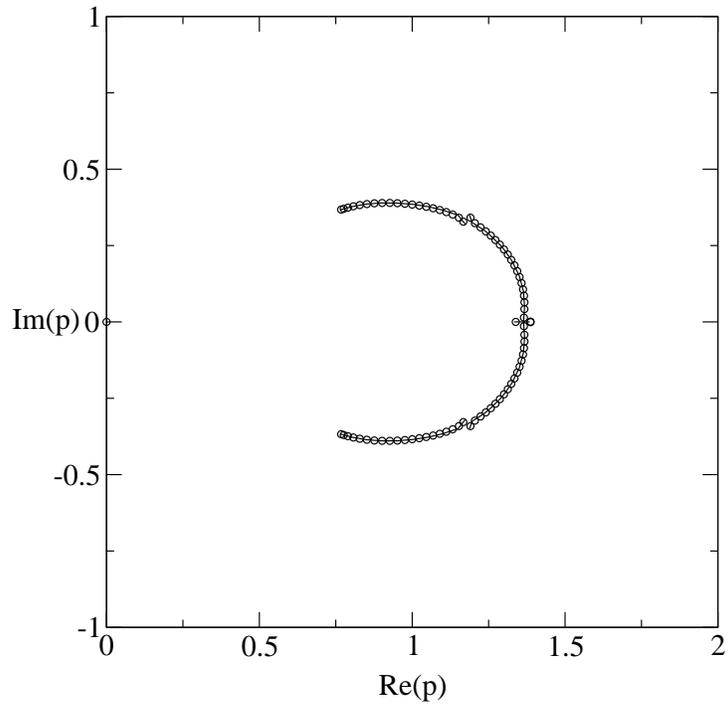}
\end{center}
\caption{\footnotesize{Singular locus ${\cal B}$ for the $L_x \to \infty$ limit
of $sqcyl(4 \times L_x)$ for either free or (twisted) periodic longitudinal
boundary conditions.  For comparison, zeros of the reliability polynomial are
shown for the cylindrical strip with $L_x=21$ (i.e., $n=84$).}}
\label{sqpy4}
\end{figure}

As before, we express the results for the $L_y=4$ strip of the square lattice
with cylindrical boundary conditions in terms of the numerator and denominator
of the generating function.  In particular, for the denominator, we find 
\beqs
{\cal D} & = & 1 + \sum_{j=1}^6 b_{sqcyl4,j}z^j \cr\cr
         & = & \prod_{j=1}^6 (1-p^4 \alpha_{sqcyl4,j}z)
\label{densqcyl4}
\eeqs
where
\beq
b_{sqcyl4,1}=-p^4(90-293p+369p^2-211p^3+46p^4)
\label{bsqcyl4_1}
\eeq
\beq
b_{sqcyl4,2}=p^8(1-p)^2(735-2977p+5094p^2-4710p^3+2481p^4-706p^5+85p^6)
\label{bsqcyl4_2}
\eeq
\beqs
b_{sqcyl4,3} & = & -p^{12}(1-p)^4(1548-7518p+15948p^2-19170p^3 \cr\cr
& & +14104p^4-6351p^5+1621p^6-181p^7)
\label{bsqcyl4_3}
\eeqs
\beq
b_{sqcyl4,4}=p^{16}(1-p)^7(735-3177p+5870p^2-5934p^3+3456p^4-1097p^5+148p^6)
\label{bsqcyl4_4}
\eeq
\beq
b_{sqcyl4,5}=-p^{20}(1-p)^{11}(90-223p+208p^2-88p^3+15p^4)
\label{bsqcyl4_5}
\eeq
\beq
b_{sqcyl4,6}=p^{24}(1-p)^{15}
\label{bsqcyl4_6}
\eeq
The degree of ${\cal D}$ is too high to obtain closed-form algebraic
expressions for the $\alpha_{sqcyl4,j}$'s, but they may be computed
numerically.  Denoting the dominant $\alpha_{sqcyl4,j}$ in the physical
interval $p \in [0,1]$ as $\alpha_{sqcyl4,max}$, one has, for the asymptotic
reliability per vertex, $r(sq[4 \times \infty,
PBC_y],p)=p(\alpha_{sqcyl4,max})^{1/4}$. In Fig. \ref{sqpy4} we show the locus
${\cal B}$ for the infinite-length limit of this strip.  This locus includes
two complex-conjugates arcs, a self-conjugate arc crossing the real axis at $p
\simeq 1.364$, and a real line segment extending from $p=4/3$ to $p \simeq
1.384$, which defines the value of $p_c$ for this strip.

\bigskip

\subsection{$L_y=2$ Strip of the Triangular Lattice with Free Boundary
Conditions}

In addition to studying how reliability polynomials and their asymptotic $r$
functions depend on the width and boundary conditions, it is also of interest
to explore how they depend on the lattice type.  For this purpose we have
carried out calculations of reliability polynomials for strips of the
triangular and honeycomb lattice.  We begin with the $L_y=2$ strip of the
triangular lattice with free boundary conditions, for which we calculate the
generating function with
\beq
{\cal N}(p,z)=p\biggl [ 1+p(1-p)z \biggr ]
\label{numgamma_txy2}
\eeq
\beqs
{\cal D}(p,z) & = & 1-p^2(7-10p+4p^2)z+p^4(1-p)^2z^2 \cr\cr
              & = & (1-p^2 \alpha_{t,2,0,1}z)(1-p^2 \alpha_{t,2,0,2}z)
\label{dengamma_txy2}
\eeqs
where
\beqs
\alpha_{t,2,0,j} & = & \frac{1}{2}\left [
7-10p+4p^2 \pm (3-2p)\sqrt{5-8p+4p^2} \right ] \cr\cr
& = & \left [ \frac{ 3-2p + \sqrt{5-8p+4p^2}}{2} \right ]^2
\quad {\rm for} \quad j=1,2 \ .
\label{alpha20j_tpxy2}
\eeqs
We note that $R(p,z)$ for the strip of the triangular lattice of length $m$ has
the factor $(8-11p+4p^2)$ if $m=1$ mod 3, where our convention is that $m=1$ is
a single square with diagonal (i.e. two triangles), and so forth.

The dominant term in the physical interval $p \in [0,1]$ is $\alpha_{t,2,0,1}$,
and hence the reliability per vertex for $L_x \to \infty$ is
\beqs
r(t[2 \times \infty, FBC_y],p) & = & p(\alpha_{t,2,0,1})^{1/2} \cr\cr
  & = & \frac{p}{\sqrt{2}} \left [ 7-10p+4p^2 \pm (3-2p)\sqrt{5-8p+4p^2}
\right ]^{1/2}
\label{rtrixy2}
\eeqs
This has derivatives
\beq
\left . \frac{dr}{dp} \right |_{p=0} = \frac{3 + \sqrt{5}}{2} = 2.61803...
\label{drdp_txy2_p0}
\eeq
and $dr/dp=0$ at $p=1$.

\begin{figure}[hbtp]
\centering
\leavevmode
\epsfxsize=4.0in
\begin{center}
\leavevmode
\epsffile{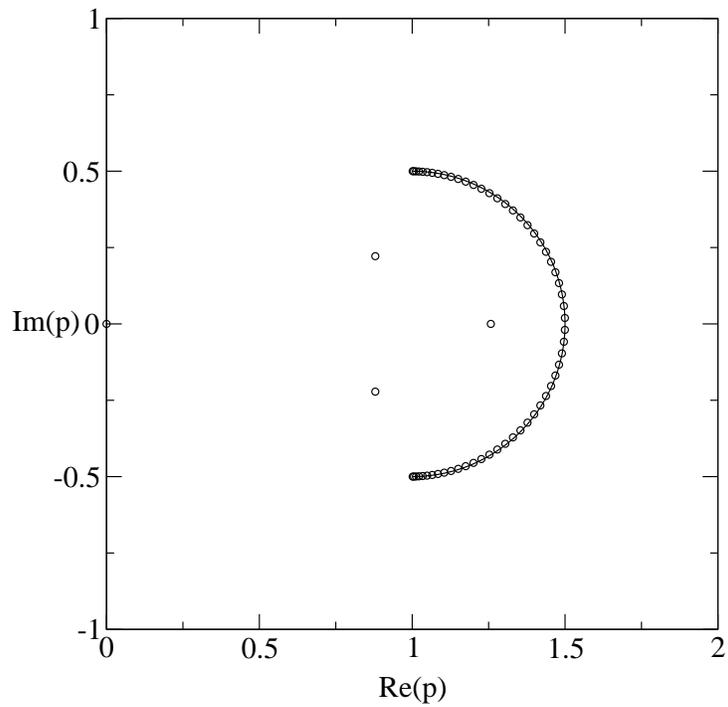}
\end{center}
\caption{\footnotesize{Singular locus ${\cal B}$ for the $L_x \to \infty$ limit
of $tri(2 \times L_x)$ for either free or (twisted) periodic longitudinal
boundary conditions.  For comparison, zeros of the reliability polynomial are
shown for the cyclic strip with $L_x=30$ (i.e., $n=60$).}}
\label{tpxy2}
\end{figure}

The locus ${\cal B}$, shown in Fig. \ref{tpxy2}, is given by an arc of a
circle:
\beq
{\cal B}: \quad p = 1 + \frac{1}{2}e^{i\theta} \ , \quad \theta \in
[-\frac{\pi}{2}, \frac{\pi}{2}]
\label{b_tpxy2}
\eeq
This is a leftwardly concave circular arc that crosses the real axis at
\beq
p_c = \frac{3}{2}  \quad  tri, \ FBC_y, \ 2 \times \infty,
\label{pctpxy2}
\eeq
and has endpoints at $p=1 \pm i/2$, where the factor $5-8p+4p^2$ in the
discriminant $(5-8p+4p^2)(3-2p)^2$ of the equation
$\alpha^2-(7-10p+4p^2)\alpha+(1-p)^2=0$ yielding the $\alpha$'s vanishes. We
shall see that a rather different strip, the self-dual strip of the square
lattice of width $L_y=1$, yields the same locus ${\cal B}$, although the
reliability polynomials are different.  Interestingly, the effective degree is
the same, namely 4, for these two strips, although the $L_y=2$ cyclic strip of
the triangular lattice is a $\Delta$-regular graph, with uniform vertex degree
$\Delta=4$, whereas the self-dual strip of the square lattice with $L_y=1$ has
two quite different kinds of vertex degrees - 3 for the outer rim and $m$ for
the central vertex connected to each vertex on the rim by edges.  However, it
is not, in general, true that if two families of graphs have the same effective
vertex degree then the resultant loci ${\cal B}$ are the same; for example,
each of the self-dual cyclic strips of the square lattice of width $L_y$ has
the same effective vertex degree in the limit $L_x \to \infty$, but they have
different loci ${\cal B}$.

\subsection{$L_y=2$ Cyclic Strip of the Triangular Lattice}

We next consider the $L_y=2$ cyclic strip of the triangular lattice.
Either using a direct calculation or as a special case of our previous
calculation of the Tutte polynomial \cite{ta}, using (\ref{rt}), we find
\beqs & & R(t[L_y=2,m,cyc],p)=p^{2m} \Biggl [
(\alpha_{t,2,0,1})^m+(\alpha_{t,2,0,2})^m - 2(\alpha_{t,2,2})^m \cr\cr & & +
m p^{-1}(5-3p)^{-1}(1-p)^2 \left \{ (\alpha_{t,2,0,1})^{m-1} (1-p) \left
(7-4p + \frac{(3-2p)(5-4p)}{\sqrt{4p^2-8p+5}} \right ) \right. \cr\cr & &
+ \left. (\alpha_{t,2,0,2})^{m-1} (1-p) \left (7-4p -
\frac{(3-2p)(5-4p)}{\sqrt{4p^2-8p+5}} \right ) -
2(p^2-3p+2)(\alpha_{t,2,2})^{m-1} \right \} \Biggr ]
\cr\cr & &
\label{Rtripxy2}
\eeqs
where $\alpha_{t,2,0,j}$ were given above for the free strip and
\beq
\alpha_{t,2,2} = (1-p)^2 \ .
\label{alpha22_tpxy2}
\eeq
Because the dominant $\alpha$'s consist of $\alpha_{t,2,0,j}$, $j=1,2$, which
are the same for the free, cyclic, and M\"obius strips, it follows that
the locus ${\cal B}$ is the same for the $L_x \to \infty$ limits of these
three strips.

\subsection{$L_y=2$ Free Strips of the Honeycomb Lattice}

For the $L_y=2$ strip of the honeycomb lattice with free boundary conditions we
calculate a generating function with
\beq
{\cal N} = p
\label{numhc}
\eeq
\beqs
{\cal D} & = & 1-p^4(6-5p)z+(1-p)p^8z^2 \cr\cr
         & = & (1-p^4 \alpha_{hc,2,0,1}z)(1-p^4 \alpha_{hc,2,0,2}z)
\label{denhc}
\eeqs
where
\beq
\alpha_{hc,2,0,j} = \frac{1}{2}\left [ 6-5p \pm \sqrt{H_2} \
\right ]
\label{alpha_hc2}
\eeq
with
\beq
H_2=32-56p+25p^2 \ .
\label{h2}
\eeq
In the physical interval $p \in [0,1]$ the dominant term is
$\alpha_{hc,2,0,1}$, so the asymptotic reliability per vertex is
\beqs
r(hc[2 \times \infty,FBC_y],p) & = & p(\alpha_{hc,2,0,1})^{1/4} \cr\cr
& = & \frac{p}{2^{1/4}} \left [6-5p + \sqrt{32-56p+25p^2} \right ]^{1/4}
\label{rhcpxy2}
\eeqs
This has derivatives
\beq
\left . \frac{dr}{dp} \right |_{p=0} = (1+\sqrt{2})^{1/2} = 1.55377...
\label{drdp_hc2_p0}
\eeq
and $dr/dp=0$ at $p=1$.

\begin{figure}[hbtp]
\centering
\leavevmode
\epsfxsize=4.0in
\begin{center}
\leavevmode
\epsffile{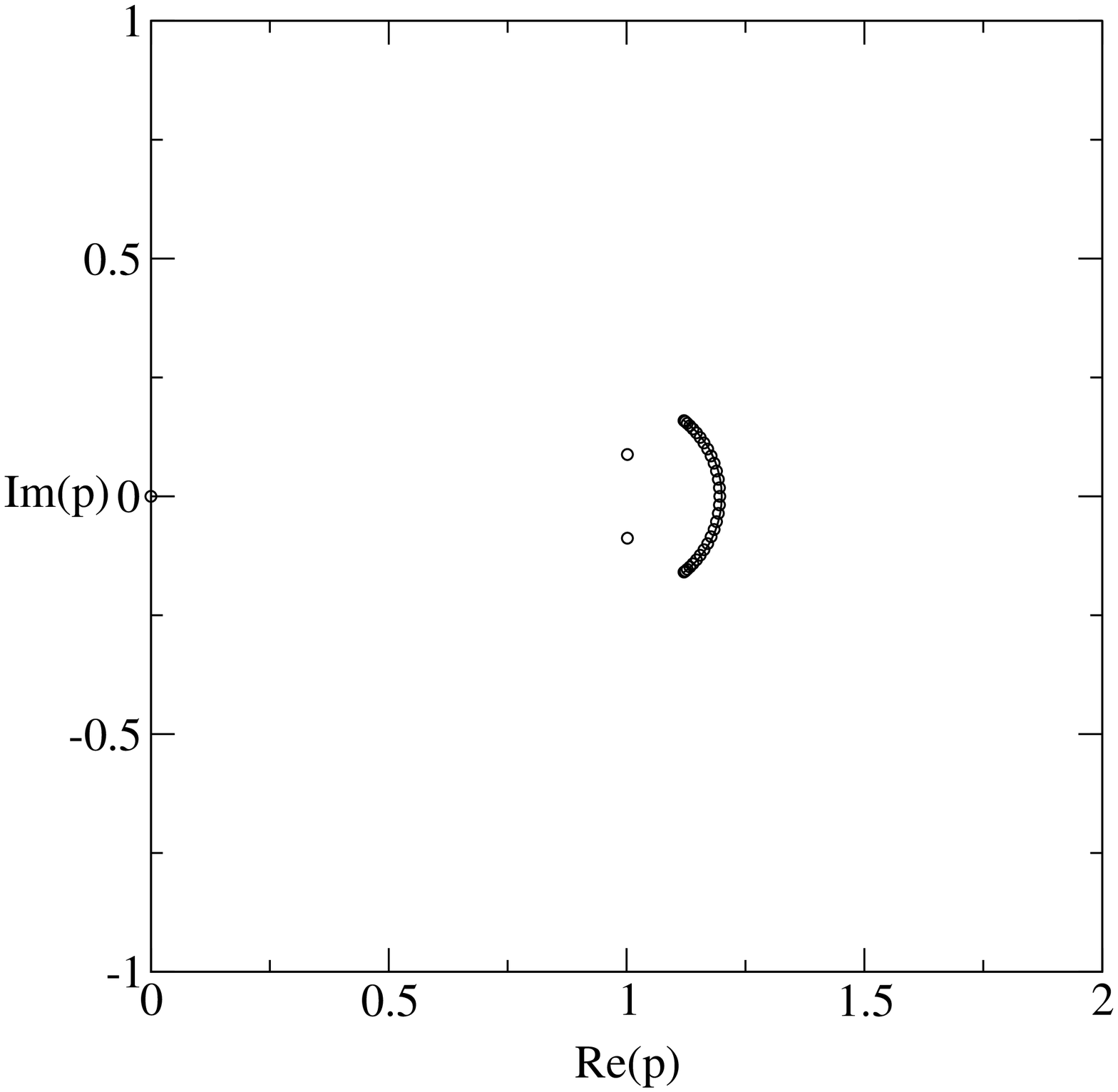}
\end{center}
\caption{\footnotesize{Singular locus ${\cal B}$ for the $L_x \to \infty$ limit
of $hc(2 \times L_x)$ for either free or (twisted) periodic longitudinal
boundary conditions.  For comparison, zeros of the reliability polynomial are
shown for the cyclic strip with $L_x=30$ (i.e., $n=120$).}}
\label{hpxy2}
\end{figure}

The locus ${\cal B}$, shown in Fig. \ref{hpxy2}, is given by an arc of a
circle:
\beq
{\cal B}: \quad p = 1 + \frac{1}{5}e^{i\theta} \ , \quad \theta \in
[-\theta_{hce}, \theta_{hce}]
\label{b_hcpxy2}
\eeq
where
\beq
\theta_{hce} = {\rm arctan}(4/3) \simeq 53.13^\circ
\label{theta_hce}
\eeq
This is a leftwardly concave arc that crosses the real axis at
\beq
p_c = \frac{6}{5}  \quad  {\rm for} \quad hc, \ FBC_y, \ 2 \times \infty,
\label{pchcpxy2}
\eeq
and has endpoints at
\beq
p_{hce}, p_{hce}^* = \frac{4}{25}(7 \pm i)
\label{phce}
\eeq
where the discriminant $H_2$ vanishes.

\subsection{$L_y=2$ Cyclic and M\"obius Strips of the Honeycomb Lattice}

For the $L_y=2$ cyclic strip of the honeycomb lattice, either using a direct
calculation or as a special case of our previous calculation of the Tutte
polynomial \cite{hca}, using (\ref{rt}), we find
\beqs
& & R(hc[L_y=2,m,cyc],p)=p^{4m} \Biggl [
(\alpha_{hc,2,0,1})^m+(\alpha_{hc,2,0,2})^m - 2(\alpha_{hc,2,2})^m \cr\cr & & +
m p^{-1}(1-p)^2 \left \{ (\alpha_{hc,2,0,1})^{m-1} \left (3 +
\frac{16-15p}{\sqrt{H_2}} \right ) \right. \cr\cr & &
\left. + (\alpha_{hc,2,0,2})^{m-1} \left (3 -
\frac{16-15p}{\sqrt{H_2}} \right ) - 2(\alpha_{hc,2,2})^{m-1}
\right \} \Biggr ] \cr\cr & &
\label{Rhcpxy2}
\eeqs
where $\alpha_{hc,2,0,j}$, $j=1,2$, were given above in (\ref{alpha_hc2}) and,
as a special case of (\ref{alphalyly}), we have
\beq
\alpha_{hc,2,2}=1-p \ .
\label{alpha_hc_22}
\eeq

For the $L_y=2$ M\"obius strip of the honeycomb lattice, using the same
methods, we obtain
\beqs
& & R(hc[L_y=2,m,Mb],p)=p^{4m} \Biggl [
(\alpha_{hc,2,0,1})^m+(\alpha_{hc,2,0,2})^m - (\alpha_{hc,2,2})^m \cr\cr & & +
m p^{-1}(1-p)^2 \left \{ (\alpha_{hc,2,0,1})^{m-1} \left (3 +
\frac{16-15p}{\sqrt{H_2}} \right ) \right. \cr\cr & &
\left. + (\alpha_{hc,2,0,2})^{m-1} \left (3 -
\frac{16-15p}{\sqrt{H_2}} \right ) + 2(\alpha_{hc,2,2})^{m-1}
\right \} \Biggr ] \ . \cr\cr & &
\label{Rhcpxy2mb}
\eeqs

Since the locus ${\cal B}$ is determined by the equality in magnitude of the
two dominant eigenvalues $\alpha_{hc,2,0,1}$ and $\alpha_{hc,2,0,2}$, and since
these are the same for the strip with free and periodic or twisted periodic
boundary conditions, it follows that this locus is the same as (\ref{b_hcpxy2})
and (\ref{theta_hce}) for the $L_y=2$ cyclic or M\"obius strip of the honeycomb
lattice.  We show this locus in Fig. \ref{hpxy2}.

\subsection{$L_y=1$ Cyclic Strip of the Square Lattice with Self-Dual Boundary
Conditions}

We have presented some general structural results above for reliability
polynomials of cyclic strip graphs of the square lattice with self-dual
boundary conditions.  Recall that these are denoted $G_D$, of size $L_y \times
L_x$; in subscripts we will use the symbol $sqdbc$. In this and the next
section we give explicit calculations of the reliability polynomials.  Note
that $G_D[1 \times L_x]$ is the wheel graph with a central spoke vertex
connected to $L_x$ outer vertices on the rim.  Either using a direct
calculation or as a special case of our previous calculation of the Tutte
polynomial \cite{jz,dg}, using (\ref{rt}), we find (with $L_x=m$)
\beq
R(G_D[1 \times m],p)=p^m\biggl [ (\alpha_{sqdbc,1,1,1})^m +
(\alpha_{sqdbc,1,1,2})^m -2(\alpha_{sqdbc,1,2,1})^m \biggr ]
\label{rwh1}
\eeq
where
\beq
\alpha_{sqdbc,1,1,j} = \frac{1}{2}\biggl [ 3-2p \pm \sqrt{5-8p+4p^2} \
\biggr ] \quad j=1,2
\label{rlam11j}
\eeq
and, consistent with (\ref{alphalyly}),
\beq
\alpha_{sqdbc,1,2,1}=1-p \ .
\label{rlam121}
\eeq
In accordance with our general results above, the total number of terms is the
same as that for the Tutte polynomial, $N_{R,G_D,L_y=1,\lambda}=
N_{T,G_D,L_y=1,\lambda}=3$.

In the physical interval $p \in [0,1]$ the dominant term is
$\alpha_{sqdbc,1,1,1}$ so that the asymptotic reliability per vertex is
\beqs
r(G_D[1 \times \infty],p) & = & p \alpha_{sqdbc,1,1,1} \cr\cr
& = & p \left [ \ \frac{3-2p + \sqrt{5-8p+4p^2}}{2} \ \right ]
\label{rgd1}
\eeqs
Interestingly, although the graphs are different and the reliability
polynomials are different for the $G_D(1 \times m)$ and $tri(2 \times m)$
strips (with either both free or both periodic longitudinal boundary
conditions), the infinite-length limits of these four families of graphs yield
the same $r$ function, i.e., (\ref{rgd1}) is identical to $r(tri[2 \times
\infty,FBC_y],p)$ given above in (\ref{rtrixy2}).

\begin{figure}[hbtp]
\centering
\leavevmode
\epsfxsize=4.0in
\begin{center}
\leavevmode
\epsffile{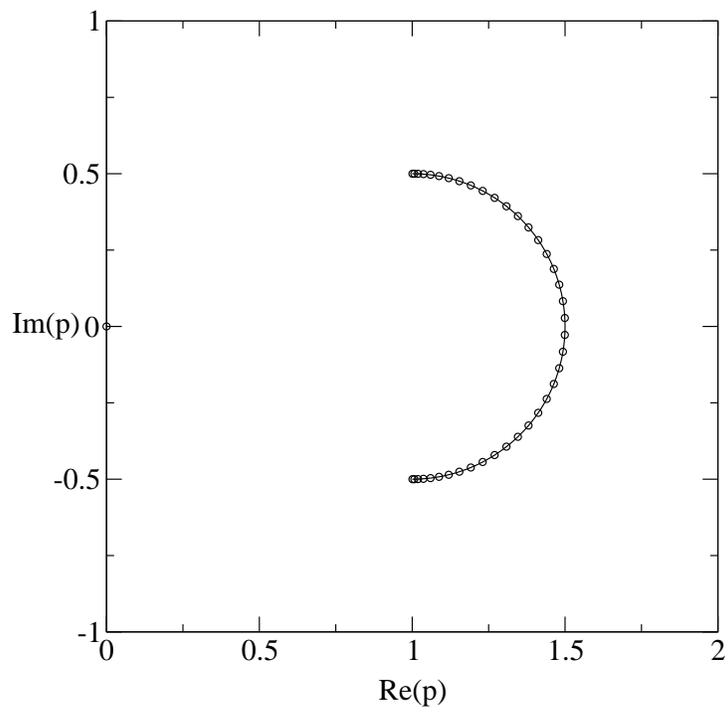}
\end{center}
\caption{\footnotesize{Singular locus ${\cal B}$ for the $L_x \to \infty$ limit
of $G_D(1 \times L_x)$. For comparison, zeros of the reliability polynomial are
shown for $L_x=40$.}}
\label{whpxy1}
\end{figure}

The locus ${\cal B}$ is shown in Fig. \ref{whpxy1} and forms arc of a circle
centered at $p=1$ of radius 1/2 with endpoints at $p=1 \pm (1/2)i$, i.e.,
\beq
{\cal B}: \quad p = 1 + \frac{1}{2}e^{i\theta} \ , \quad \theta \in
[-\frac{\pi}{2}, \frac{\pi}{2}]
\label{b_whpxy1}
\eeq
so that
\beq
p_c = \frac{3}{2} \quad {\rm for} \quad G_D(1 \times \infty) \ .
\label{pcwhpxy1}
\eeq
The endpoints occur at the branch point singularities of the polynomial
$5-8p+4p^2$ that appears in the square root in the terms (\ref{rlam11j}).
The locus (\ref{b_whpxy1}) is identical to (\ref{b_tpxy2}), although the
$\alpha$'s involved are different.

\bigskip

\subsection{$L_y=2$ Cyclic Strip of the Square Lattice with Self-Dual Boundary
Conditions}

For $G_D[2 \times m]$, our general formulas give $N_{R,G_D,2,\lambda}=10$ with
$n_R(G_D,2,1)=5$, $n_R(G_D,2,2)=4$, and $n_R(G_D,2,3)=1$.  We find that
$R(G_D[2 \times m],p)$ has the form (\ref{rgd}),
\beq
R(G_D[2 \times m],p)=p^{2m}\biggl [
\sum_{j=1}^5 (\alpha_{sqdbc,2,1,j})^m -2 \sum_{j=1}^4
(\alpha_{sqdbc,2,2,j})^m + 3(\alpha_{sqdbc,2,3,1})^m \biggr ]
\label{rwh2}
\eeq
where
\beq
\alpha_{sqdbc,2,3,1}=(1-p)^2
\label{lam_sqdbc2_231}
\eeq
and the $\alpha_{sqdbc,2,2,j}$ are the roots of the degree-4 equation
\beq
\xi^4+a_{db1}\xi^3+a_{db2}\xi^2+a_{db3}\xi+a_{db4}=0
\label{dbcly2deg4eq}
\eeq
with
\beq
a_{db1}=-(1-p)(7-5p)
\label{adb1}
\eeq
\beq
a_{db2}=(1-p)^2(13-17p+5p^2)
\label{adb2}
\eeq
\beq
a_{db3}=-(1-p)^4(7-5p)
\label{adb3}
\eeq
\beq
a_{db4}=(1-p)^6 \ .
\label{adb4}
\eeq
In eq. (\ref{rwh2}) the $\alpha_{sqdbc,2,1,j}$ are the roots of the
degree-5 equation
\beq
\xi^5+b_{db1}\xi^4+b_{db2}\xi^3+b_{db3}\xi^2+b_{db4}\xi+b_{db5}=0
\label{dbcly2deg5eq}
\eeq
where
\beq
b_{db1}=-(12-18p+7p^2)
\label{bdbc1coeff}
\eeq
\beq
b_{db2}=(1-p)(36-67p+41p^2-8p^3)
\label{bdbc2coeff}
\eeq
\beq
b_{db3}=-(1-p)^2(36-75p+53p^2-13p^3)
\label{bdbc3coeff}
\eeq
\beq
b_{db4}=2(1-p)^4(3-2p)(2-p)
\label{bdbc4coeff}
\eeq
\beq
b_{db5}=-(1-p)^6 \ .
\label{bdbc5coeff}
\eeq

\begin{figure}[hbtp]
\centering
\leavevmode
\epsfxsize=4.0in
\begin{center}
\leavevmode
\epsffile{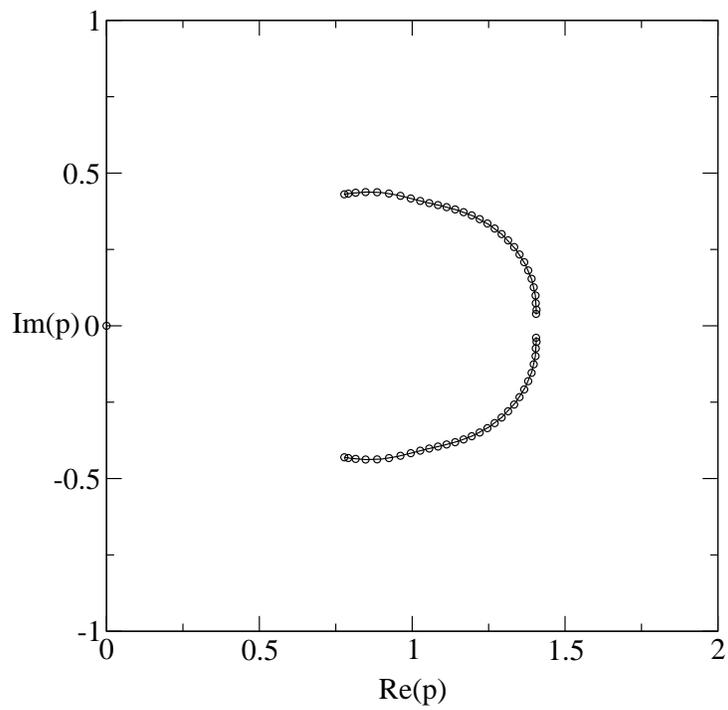}
\end{center}
\caption{\footnotesize{Singular locus ${\cal B}$ for the $L_x \to \infty$ limit
of $G_D(2 \times L_x)$. For comparison, zeros of the reliability polynomial are
shown for $L_x=30$ (i.e., $n=61$).}}
\label{whpxy2}
\end{figure}

The locus ${\cal B}$ is shown in Fig. \ref{whpxy2} and consists of arcs, again
concave to the left, that almost cross, but actually have endpoints near to,
the real axis at $q \simeq 1.4$ and have endpoints at $q \simeq 0.7765 \pm
0.4302i$ and $q \simeq 1.406 \pm 0.036795i$.

\subsection{$L_y=3$ Cyclic Strip of the Square Lattice with Self-Dual Boundary
Conditions}

\begin{figure}[hbtp]
\centering
\leavevmode
\epsfxsize=4.0in
\begin{center}
\leavevmode
\epsffile{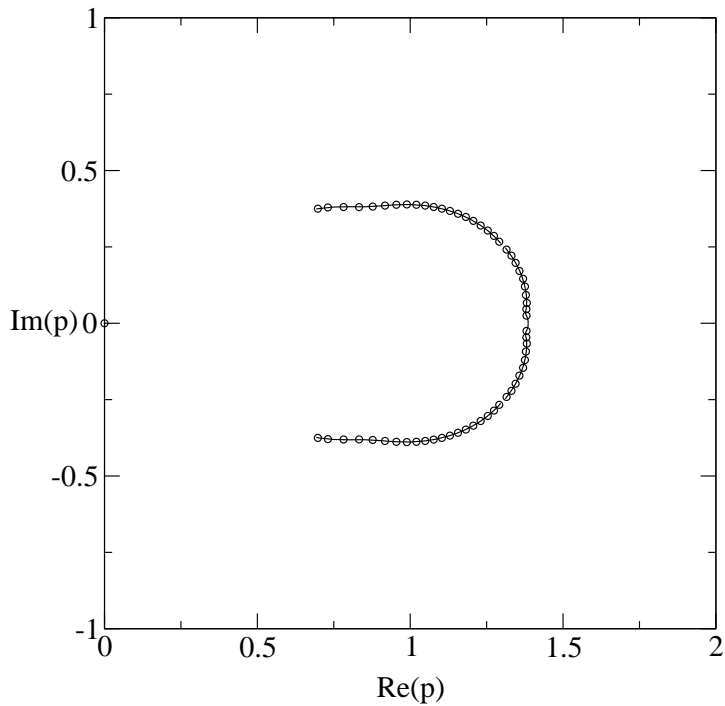}
\end{center}
\caption{\footnotesize{Singular locus ${\cal B}$ for the $L_x \to \infty$ limit
of $G_D(3 \times L_x)$.  For comparison, zeros of the reliability polynomial
are shown for $L_x=20$ (i.e., $n=61$).}}
\label{whpxy3}
\end{figure}

For $G_D[3 \times m]$, our general formulas give $N_{R,G_D,3,\lambda}=35$ with
$n_R(G_D,3,4)=1$, $n_R(G_D,3,3)=6$, $n_R(G_D,3,2)=14$, and $n_R(G_D,3,1)=14$.
We have calculated the reliability polynomial from our earlier calculation of
the Tutte polynomial for this strip \cite{dg,sdg}.  Since the results for the
$\alpha$'s are rather complicated, we do not list them in detail here (some
examples are given in the appendix) but instead concentrate on the locus
${\cal B}$.  This is shown in Fig. \ref{whpxy3}.

\subsection{Some Families of Graphs with Multiple Edges}

Consider a connected graph $G=(V,E)$.  We take this graph to be loopless; this
incurs no loss of generality since loops do not affect the reliability
polynomial.  Now consider a graph $H$ obtained from $G$ by adding another edge.
Clearly, for the physical range $p \in [0,1]$, the inequality $R(H,p) \ge
R(G,p)$ holds, with equality only at $p=1$.  That is, adding further
communications link(s) between two (already connected) nodes in a network
increases the reliability of the network.  A particularly simple modification
in the graph is to replace each edge joining two vertices by $\ell$ edges
joining these edges, thereby obtaining what we have denoted $G_\ell$ above.
Given any such graph $G$, whether or not it is a member of a recursive family,
our theorem above with (\ref{rgell}) and (\ref{pprime}) enables one to
calculate $R(G_\ell,p)$.  Here we shall present some illustrations of this.
For definiteness, we consider the $L_y=2$ cyclic strip of the square lattice.
Using our calculation of the reliability polynomial for the $L_y=2$ strip of
the square lattice together with our theorem, we have calculated the
reliability polynomial for the same strip with all edges replaced by
$\ell$-fold multiple edges joining the same vertices.

We have calculated the loci ${\cal B}$ in the $L_x \to \infty$ limit for these
strips.  The single point at $p=4/3$ where the locus ${\cal B}$ crosses the
real $p$ axis for the $sq[2 \times \infty]$ case, shown in Fig. \ref{sqpxy2},
corresponds to $\ell$ separate points for the $\ell$-fold edge replication of
$sq[2 \times \infty]$.  These can be calculated using (\ref{rgell}) with
(\ref{pprime}), and we obtain
\beq
p = 1 - \frac{1}{3^{1/\ell}}e^{\frac{i\pi (1+2k)}{\ell}} \ , \quad
k=0,1,...,\ell-1
\label{pcross_ell}
\eeq
For example, for $\ell=2$, this yields the two values $p=1 \pm i/\sqrt{3}$, and
so forth for higher values of $\ell$.  The arc endpoints can be calculated in
the same way.

It is also of interest to study strip graphs in which a certain subset of the
edges are replaced by $\ell$-fold replicated edges joining the same vertices.
A simple example of this is already provided by the $L_y=2$ strips of the
square lattice with cylindrical, torus, or Klein bottle boundary conditions,
which have double transverse edges.  We have calculated the reliability
polynomials for the cyclic $L_y=2$ strips of the square lattice with
$\ell$-fold multiple transverse edges.  We denote these strip graphs as
$sq[2 \times m, cyc., v\ell]$.  We find
\beqs
& & R(sq[L_y=2,m,cyc,v\ell],p)=p^{2m} \Biggl [
(\alpha_{sqcyc2,v\ell,1})^m+(\alpha_{sqcyc2,v\ell,2})^m -
2(\alpha_{sqcyc2,v\ell,3})^m  \cr\cr & & +
m p^{-1}(1-p)^{\ell+1} \left \{ (\alpha_{sqcyc2,v\ell,1})^{m-1} \left (
VL1(p,\ell) + VL2(p,\ell) \right ) \right. \cr\cr & &
\left. + (\alpha_{sqcyc2,v\ell,2})^{m-1} \left ( VL1(p,\ell) - VL2(p,\ell)
\right ) - (\alpha_{sqcyc2,v\ell,3})^{m-1} \right \} \Biggr ] \cr\cr & &
\label{Rsqpxy2vl}
\eeqs
where
\beqs
& & \alpha_{sqcyc2,v\ell,j} = (2p)^{-1}\Biggl [ 2-p-(2-3p)(1-p)^\ell \cr\cr
& & \pm [(1-(1-p)^\ell)((2-p)^2-(2-3p)^2(1-p)^\ell)]^{1/2} \Biggr ],
\quad j=1,2
\label{sqcycvl12}
\eeqs
\beq
\alpha_{sqcyc2,v\ell,3}=(1-p)^\ell
\label{sqcycvl3}
\eeq
and
\beq
VL1(p,\ell)=\frac{(1-p)^{\ell-1}(2p-1)+1}{2p(1-p)^{\ell-1}}
\label{vl1}
\eeq
\beq
VL2(p,\ell)=\frac{(2p-1)(3p-2)(1-p)^\ell-4+10p-7p^2
+\frac{2-p}{(1-p)^{\ell-1} } }
{2p [(1-(1-p)^\ell)((2-p)^2-(2-3p)^2(1-p)^\ell)]^{1/2}}
\label{vl2}
\eeq

\begin{figure}[hbtp]
\centering
\leavevmode
\epsfxsize=4.0in
\begin{center}
\leavevmode
\epsffile{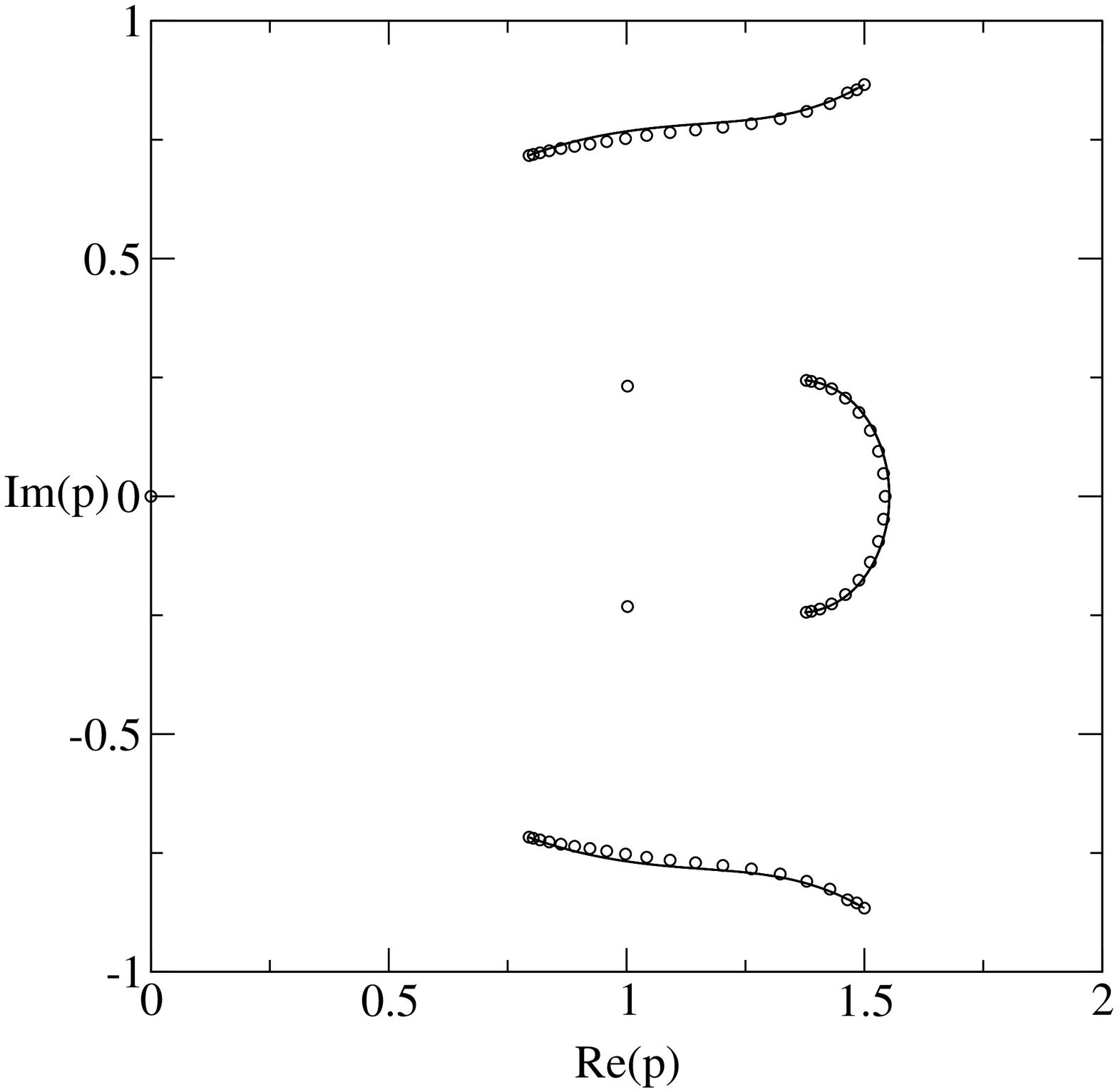}
\end{center}
\caption{\footnotesize{Singular locus ${\cal B}$ for the $L_x \to \infty$ limit
of the $L_y=2$ strip of the square lattice with $\ell=3$-fold multiple
transverse edges.  For comparison, zeros of the reliability polynomial are
shown for this strip with cyclic boundary conditions and $L_x=20$.}}
\label{sqpxy2v3}
\end{figure}

\begin{figure}[hbtp]
\centering
\leavevmode
\epsfxsize=4.0in
\begin{center}
\leavevmode
\epsffile{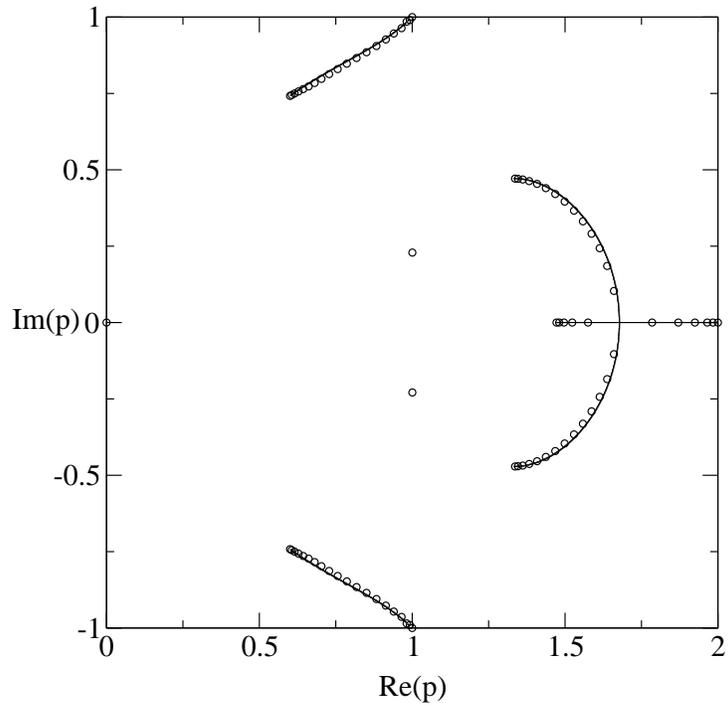}
\end{center}
\caption{\footnotesize{Singular locus ${\cal B}$ for the $L_x \to \infty$ limit
of the $L_y=2$ strip of the square lattice with $\ell=4$-fold multiple
transverse edges. For comparison, zeros of the reliability polynomial
are shown for this strip with cyclic boundary conditions and $L_x=20$.}}
\label{sqpxy2v4}
\end{figure}

\bigskip

In Figs. \ref{sqpxy2v3} and \ref{sqpxy2v4} we show the loci ${\cal B}$ in the
limit $L_x \to \infty$ for the cases $\ell=3,4$.  In general, ${\cal B}$
contains arcs protruding inward from points on the circle $|p-1|=1$ given by
$p=1-\exp(2i \pi k/\ell)$ (cf. eq. (\ref{pzero_t2})) for $0 \le k \le \ell-1$.
If and only if $\ell$ is even, then one of these points, namely the one
corresponding to $k=\ell/2$, occurs at $p=2$, and the part of ${\cal B}$ that
connects with this point is a line segment on the real axis extending from
$p=2$ downward to an $\ell$-dependent lower limit.

\subsection{$L_y=2$ Free Strip of the $sq_d$ Lattice with Multiple Transverse
Edges}

\begin{figure}[hbtp]
\centering
\leavevmode
\epsfxsize=4.0in
\begin{center}
\leavevmode
\epsffile{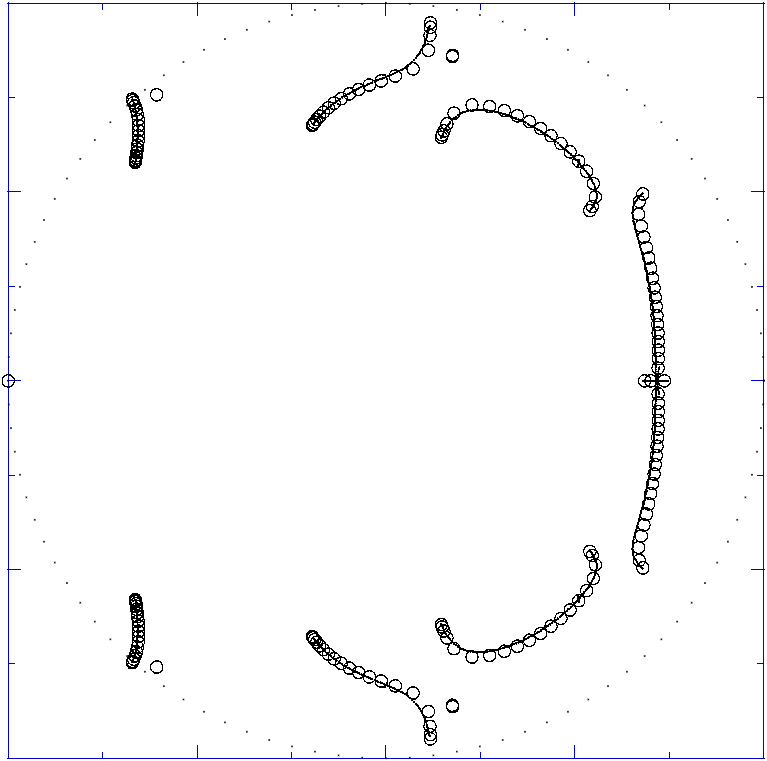}
\end{center}
\caption{\footnotesize{Singular locus ${\cal B}$ for the $L_x \to
\infty$ limit of the $L_y=2$ strip of the $sq_d$ lattice with
6-fold multiple transverse edges. For comparison, zeros of the
reliability polynomial are shown for this strip with free boundary
conditions and $L_x=21$. Horizontal and vertical axes are $Re(p)$ and $Im(p)$
with tick spacing 0.25, and the circle $|p-1|=1$ is shown with light dots.}}
\label{k4y2v6k}
\end{figure}

In this section we exhibit the first of several recursive families of lattice
strip graphs that we have found with the property that some zeros of their
reliability polynomials and, in the limit of infinite length, some portions of
their accumulation sets ${\cal B}$, lie outside the disk $|p-1|$.  This family
is a one-parameter family generalizing the lowest-order graph found by Royle
and Sokal.  That graph is constructed as follows \cite{bcf}: start with the
complete graph $K_4$ and choose two nonintersecting edges; replace each of
these edges by six edges in parallel, i.e., joining the same pair of vertices.
(Here the complete graph $K_n$ is defined as the graph with $n$ vertices such
that every vertex is connected by one edge to every other vertex.)  We had
previously calculated Tutte and chromatic polynomials for strips composed of
$K_4$ subgraphs \cite{w3,k,ka}.  One of the equivalent ways of defining these
strip graphs is to start with a strip of the square lattice and add edges
connecting diagonally opposite vertices of each square, thereby replacing each
square by a $K_4$; they were thus denoted strips of the $sq_d$ lattice, $sq_d(2
\times L_x, BC_x)$, where $BC_x$ indicated the longitudinal boundary
conditions.  In a physical context, the Potts model on the $sq_d$ lattice is
equivalent to the Potts model on the square lattice with next-nearest-neighbor
spin-spin couplings.  The longitudinal ($x$) and transverse ($y$) directions on
this strip are taken to be horizontal and vertical, respectively.  In
particular, we consider a $sq_d$ strip with width $L_y=2$ and arbitrary length
$L_x$.  We then replace each vertical edge by six edges joining the same pair
of vertices (leaving the horizontal and diagonal edges unchanged). An
elementary proof \cite{k} shows that for free longitudinal boundary conditions
this is a planar graph.  We calculate the reliability polynomial and find for
the numerator and denominator of the generating function
\beq
{\cal N}(p,z)=p\biggl [ A_0 + A_1 z \biggr ]
\label{numgamma_sqdv6}
\eeq
where
\beq
A_0 = -(p-2)(p^2-3p+3)(p^2-p+1)
\label{a0sqdv6}
\eeq
\beq
A_1 = -p^2(p-1)^6(3p-4)
\label{a1sqdv6}
\eeq
\beq
{\cal D}(p,z)=1 + b_1z + b_2z^2
\label{densqdv6}
\eeq
where
\beq
b_1 = -p^2(6p^8-52p^7+200p^6-448p^5+644p^4-616p^3+391p^2-156p+32)
\label{b1sqdv6}
\eeq
\beq
b_2 = -2p^5(p-2)(p-1)^8
\label{b2sqdv6}
\eeq
Using this exact calculation, one easily verifies that members of this family
have reliability polynomials with zeros outside the disk $|p-1| \le 1$.  The
locus ${\cal B}$ for the infinite-length limit of this family is shown in
Fig. \ref{k4y2v6k}. As is evident, ${\cal B}$ consists of arcs, and the tips of
two complex-conjugate arcs on the upper and lower left end at the points
\beq
p = 0.327752 \pm 0.747464i
\label{k4y2v6k_endpoint}
\eeq
which have
\beq
|p-1| = 1.005296
\label{k4y2v6kviolation}
\eeq
These arc endpoints and their locally neighboring sections of their arcs thus
lie outside of the disk $|p-1| \le 1$.  Since ${\cal B}$ is the continuous
accumulation set of the zeros of the reliability polynomial, it follows that as
$L_x \to \infty$, infinitely many zeros of the reliability polynomial lie
outside this disk.  Interestingly, although the locus ${\cal B}$ does extend
outside the disk $|p-1| \le 1$, the amount by which it does so is quite small,
as is evident from the positions of the arc endpoints given in eq. 
(\ref{k4y2v6kviolation}).  This property is also true of the other recursive 
families that we have found with zeros of $R(G,p)$ and limiting loci 
${\cal B}$ lying outside the disk $|p-1| \le 1$.  

\begin{figure}[hbtp]
\centering
\leavevmode
\epsfxsize=4.0in
\begin{center}
\leavevmode
\epsffile{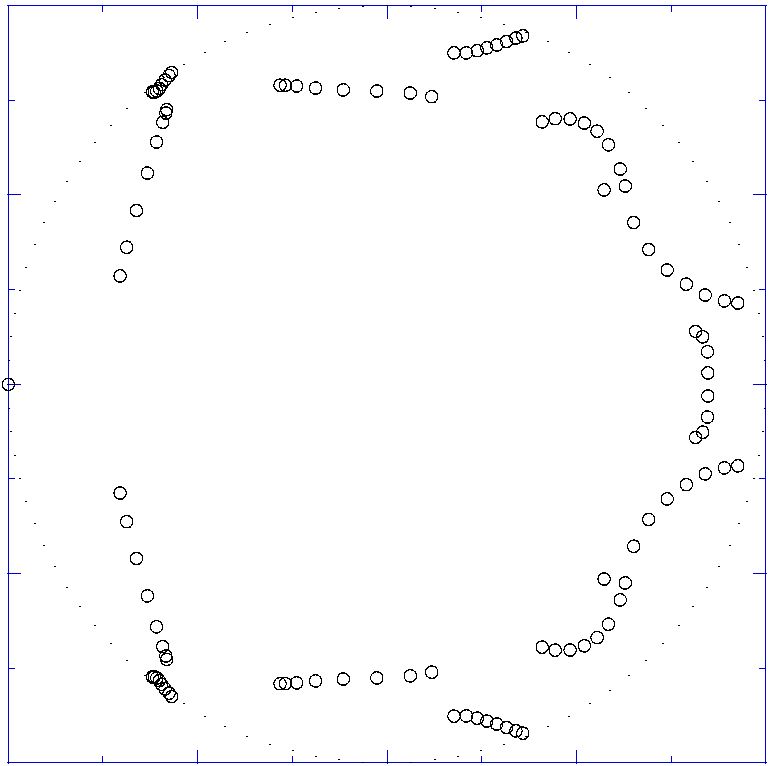}
\end{center}
\caption{\footnotesize{Zeros of the reliability polynomial for the
$L_y=2,L_x=9$ strip of the $sq_d$ lattice with 6-fold multiple longitudinal
edges and free boundary conditions.  Horizontal and vertical axes are $Re(p)$
and $Im(p)$ with tick spacing 0.25, and the circle $|p-1|=1$ is shown with 
light dots.}}
\label{k4y2h6k}
\end{figure}

We have also calculated the reliability polynomial for the analogous strip
graphs with free longitudinal boundary conditions and with each longitudinal,
rather than transverse, edge replaced by 6 edges joining the same pair of
vertices.  We find similar results. A plot of the zeros for $L_x=9$ is shown in
Fig. \ref{k4y2h6k}.  The arcs on ${\cal B}$ that extend outside the disk
$|p-1| \le 1$ are again situated in the upper and lower left, and the endpoints
of these arcs that extend outside this disk are
\beq
p = 0.4346475 \pm 0.8266808i
\label{k4y2h6k_endpoint}
\eeq
with
\beq
|p-1| = 1.001511
\label{sqdh6_violation}
\eeq

\subsection{$L_y=2$ Free Strip of $K_4$ Subgraphs with Multiple Edges}

\begin{figure}[hbtp]
\centering \leavevmode \epsfxsize=4.0in
\begin{center}
\leavevmode \epsffile{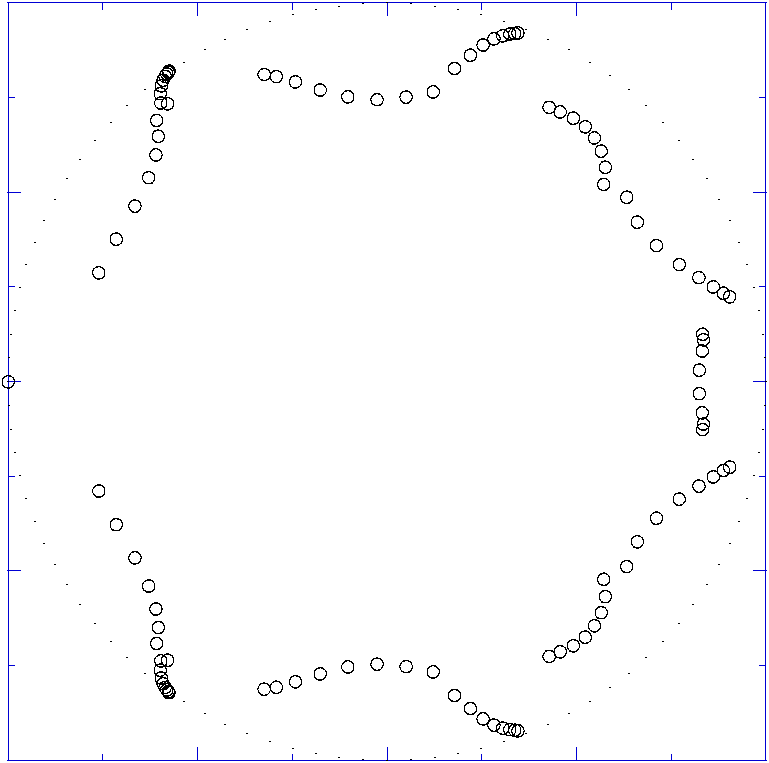}
\end{center}
\caption{\footnotesize{Zeros of the reliability polynomial for the
$L_y=2, L_x=9$ free strip of $K_4$ subgraphs with 6-fold multiple
longitudinal edges. Horizontal and vertical axes are $Re(p)$
and $Im(p)$ with tick spacing 0.25, and the circle $|p-1|=1$ is shown with
light dots.}}
\label{k4k2planark}
\end{figure}

We consider the strip shown in Fig. 8 of the Appendix of \cite{k}, but with
free, instead of cyclic, longitudinal boundary conditions.  In the notation of
that figure, we replace each of the edges (labelled as $e_{ij}=(v_i,v_j)$)
(1,2), (2,3), etc. on the top; (4,5), (5,6), etc. on the bottom, and (1,7),
(5,8), (2,9), etc. in the middle with a 6-fold replicated edge.  This is a
family of planar graphs (as is the family in \cite{k} with cyclic boundary
conditions).  We calculate the reliability polynomial and find for the
numerator and denominator of the generating function
\beq
{\cal N}(p,z)=p\biggl [ A_0 + A_1 z \biggr ]
\label{numgamma_sqd_appendix}
\eeq
where
\beq
A_0 = 1
\label{k4k2e6_appendix0}
\eeq
\beq
A_1= p^2(p-1)(p-2)(p^2-p+1)(p^2-3p+3)(2p^6-14p^5+42p^4-70p^3+70p^2-42p+13)
\label{k4k2e6_appendix1}
\eeq
\beq
{\cal D}(p,z)=1 + b_1z + b_2z^2
\label{densqd_appendix}
\eeq
where
\beqs
& & b_1 = p^2(6p^{13}-86p^{12}+574p^{11}-2366p^{10}+6734p^9-14014p^8+22018p^7
\cr\cr
& & -26561p^6+24731p^5-17661p^4+9471p^3-3647p^2+918p-118)
\label{b1_k4k2e6_appendix6}
\eeqs
\beqs & & b_2 =
p^4(1-p)^3(4p^{22}-100p^{21}+1201p^{20}-9224p^{19}+50876p^{18}-214544p^{17}
\cr\cr & &
+719020p^{16}-1965184p^{15}+4459758p^{14}-8511200p^{13}+13781747p^{12}-19045712p^{11}
\cr\cr & &
+22536020p^{10}-22849416p^9+19815434p^8-14628596p^7+9116492p^6
\cr\cr & & -4732392p^5+2004288p^4-670184p^3+167360p^2-28048p+2401)
\label{b2_k4k2e6_appendix6} \eeqs
Zeros of the reliability polynomial for the $L_x=9$ member of this family are
shown in Fig. \ref{k4k2planark}. The tips of two complex-conjugate arcs, again
on the upper and lower left, extend outside the disk $|p-1| \le 1$,
ending at the points
\beq
p = 0.4254334 \pm 0.8216255 i
\label{k4k2_appendix6_endpoint}
\eeq
which have
\beq
|p-1| = 1.002594
\label{k4k2_appendix6_violation}
\eeq
Thus, as $L_x \to \infty$, an infinite number of zeros of the reliability
polynomial accumulate to form part of the locus ${\cal B}$ which extends
outside the disk $|p-1| \le 1$.  As before, the amount by which some zeros and
${\cal B}$ lie outside the disk is small.

\section{Complete Graphs}

\begin{figure}[hbtp]
\centering
\leavevmode
\epsfxsize=4.0in
\begin{center}
\leavevmode
\epsffile{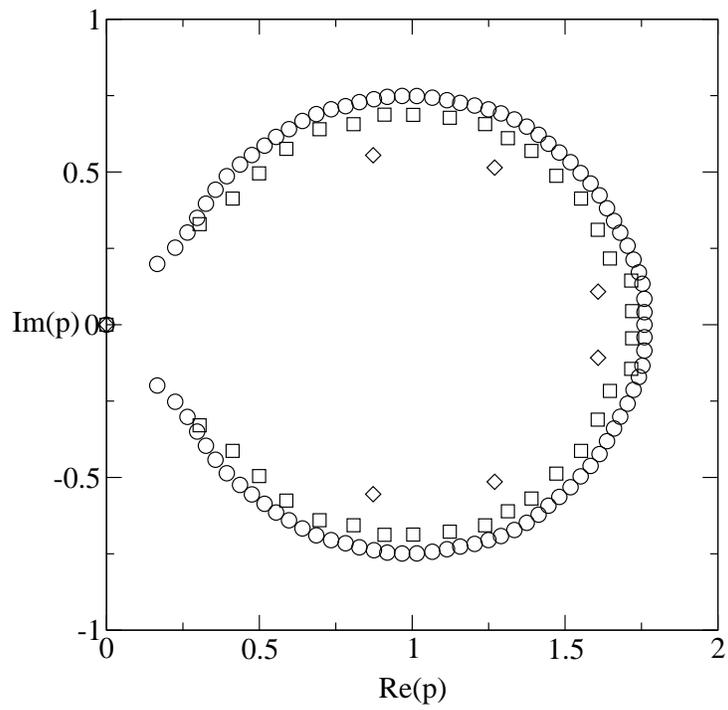}
\end{center}
\caption{\footnotesize{Plot of zeros of the reliability polynomial
$R(K_n,p)$ for $n=$ (a) 5 \ ($\Diamond$), (b) 10 \ ($\Box$), (c) 15 \
($\circ$).}}
\label{complete}
\end{figure}

The complete graph $K_n$ is the graph with $n$ vertices such that every vertex
is connected by edges to every other vertex.  The family of complete graphs is
not a recursive family, and provides a contrast to the recursive families on
which we have concentrated in this paper.  The reliability polynomial for $K_n$
is \cite{colbourn}
\beq
R(K_n,p) = 1 - \sum_{j=1}^{n-1} {n-1 \choose j-1} (1-p)^{j(n-j)}R(K_j,p) \ .
\label{rkn}
\eeq
Since this family is not recursive, the reliability polynomial does not have
the form (\ref{rgsum}) and our usual methods for determining an asymptotic
accumulation set of zeros as the solution of the degeneracy in magnitudes of
dominant $\alpha$ terms do not apply.  However, we have studied the zeros of
$R(K_n,p)$ for a wide range of values of $n$, and we find that these zeros
typically form an oval-shaped pattern that is concave to the left and is
roughly centered around $p=1$, somewhat similar to the exact results that we
have established for our recursive families of graphs without multiple edges.
As illustrations, we show in Fig. \ref{complete} the patterns of zeros for
$n=5,10$, and 15.  We observe that as $n$ increases, the oval moves outward
from $p=1$.  Note that since $K_n$ is a $\Delta$-regular graph with
$\Delta=n-1$, it follows that the vertex degree $\Delta \to \infty$ as $n \to
\infty$.  This feature that $\Delta \to \infty$ as $n \to \infty$ is thus
shared in common by the two families $K_n$ and the thick links $TL_n$.  For the
latter family, it is elementary that as $n \to \infty$, ${\cal B}$ is the
circle $|p-1|=1$.  However, our results show that it is not necessary for any
vertex degree to go to infinity in order for a part of ${\cal B}$ to satisfy
$|p-1| \ge 1$.  Our results for the family of wheel graphs, $G_D(1 \times m)$
also show that the property that a vertex degree goes to infinity as $|V| \to
\infty$ is not sufficient for ${\cal B}$ to have a part with $|p-1| \ge 1$.

\section{Discussion of Results}

In this section we discuss some general features of our results.

\begin{figure}[hbtp]
\centering \leavevmode \epsfxsize=4.0in
\begin{center}
\leavevmode
\epsffile{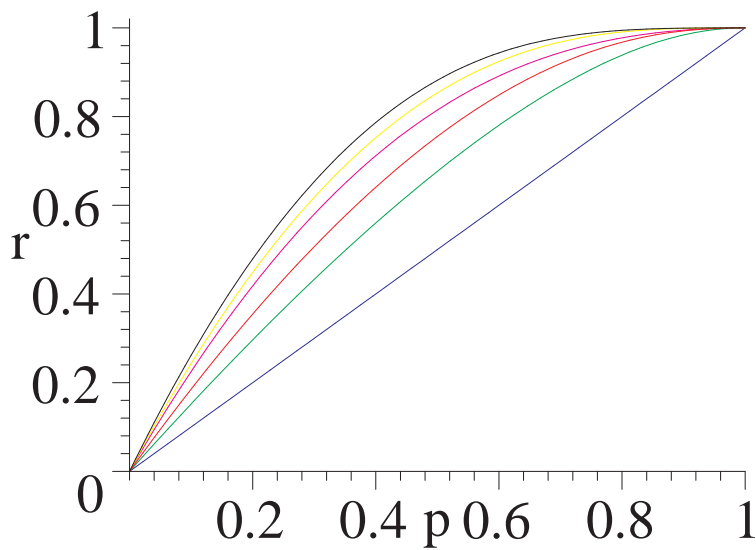}
\end{center}
\caption{\footnotesize{Comparison of asymptotic reliability per vertex $r$ for
the $L_x \to \infty$ limit of several lattice strips.  From bottom to top, the
curves refer to
(a) $sq[1 \times \infty,FBC_y]$ with $d_{eff}=2$,
(b) $hc[2 \times \infty,FBC_y]$ with $d_{eff}=2.5$,
(c) $sq[2 \times \infty,FBC_y]$ with $d_{eff}=3$,
(d) $sq[2 \times \infty,PBC_y]$ with $d_{eff}=4$,
(e) $G_D[1 \times \infty]$, same $r$ as for
$tri[2 \times \infty]$, with $d_{eff}=4$,
(f) $G_D[2 \times \infty]$, with $d_{eff}=4$.}}
\label{rcomparison}
\end{figure}

\begin{itemize}

\item

For a given type of lattice and a given width and choice of transverse
boundary conditions, we find that in the infinite-length limit, the reliability
per vertex $r$ is the same for any choice of longitudinal boundary conditions.

\item

On general grounds, one expects that the higher the connectivity, and,
equivalently for our present purposes, the higher the effective vertex degree
$d_{eff}$, the larger the value of $r$ for a fixed $p$.  In
Fig. \ref{rcomparison} we plot $r$ for some strips of the square, triangular,
and honeycomb lattices.  We see that our expectation is confirmed by our exact
results.

\item

Closely related to this, our results exhibit the property that the derivative
$dr/dp$ at $p=0$ is an increasing function of $d_{eff}$.  Except for the
$L_y=1$ strips of the square lattice, i.e., the line and circuit graphs, we
find that $dr/dp=0$ at $p=1$ for all strips that we have studied.  In Table
\ref{properties} we list some of the properties that we have found from our
calculations.

\item

We can also comment on features of the zeros of the reliability polynomials and
their continuous accumulation sets ${\cal B}$.  We find that the zeros of the
reliability polynomials for many strips studied do exhibit the property that
$|p-1| \le 1$, as do their loci ${\cal B}$ in the limit of infinite length.
However, motivated by \cite{bcf}, we have calculated reliability polynomials
for strips such that the zeros of $R(G,p)$ violate the bound $|p-1| \le 1$ and
such that, in the limit of infinite length, the continuous accumulation sets of
zeros ${\cal B}$ extend outside of the disk $|p-1| \le 1$.  An intriguing
result is that the amounts by which some individual zeros, and the outer part
of the loci ${\cal B}$, lie outside this disk are small.

\item

We find that in some cases, ${\cal B}$ crosses the real axis and hence defines
a $p_c$, but in other cases, it does not.  However, typically, even for
families for which ${\cal B}$ does not cross the real axis, there are arc
endpoints on ${\cal B}$ that lie close to this axis, thereby enabling one
to define a $(p_c)_{eff}$ via extrapolation or just as the real part of the
nearby endpoints.  Our results have the property that these values of $p_c$ or
$(p_c)_{eff}$ lie in the interval $p \in (1,2]$.  It is interesting to discuss
the relation between this result and the Brown-Colbourn theorem \cite{bc} that
$R(G,p)$ does not have any real zeros in the region outside of the set $0 \cup
(1,2]$.  In general, the property that a polynomial has no zeros in an interval
of the real axis does not exclude the possibility that as the degree of this
polynomial goes to infinity, a resultant continuous accumulation set of zeros
${\cal B}$ crosses the real axis in this interval.  A well-known case where
this phenomenon does occur is provided by the Potts model partition function.
This is a polynomial in $q$ and $a=e^K$ with non-negative coefficients.  Thus,
for fixed positive $q$, for either sign of the spin-spin coupling $J$, this
partition function cannot have any zeros on the positive real $a$ axis.
However, for the infinite limit of the square lattice, the locus ${\cal B}$ in
this case does cross the positive real $a$ axis at the point $a_c =
1+\sqrt{q}$, which serves as the phase boundary between the paramagnetic and
ferromagnetic phases occupying the respective intervals $1 \le a \le a_c$ and
$a_c \le a \le \infty$.

\item

For all of the lattice strip graphs for which we have calculated $R(G,p)$ and
${\cal B}$, we find that ${\cal B}$ is independent of the longitudinal boundary
conditions, although it depends on the transverse boundary conditions and the
type of lattice.  This requires that the dominant $\alpha$ in the case of
cyclic strips must arise from the subset of the $\alpha$'s with coefficient
$c^{(0)}$ that coincides with the set of $\alpha$'s for the corresponding strip
with free longitudinal boundary conditions.  The property that, for a given
type of lattice strip, the locus ${\cal B}$ is independent of the longitudinal
boundary conditions is quite different from what we found for the corresponding
continuous accumulation sets of zeros of chromatic polynomials of lattice
strips \cite{bcc}, \cite{w}-\cite{k}, which do depend on both longitudinal and
transverse boundary conditions.

\item

For all of the lattice strip graphs for which we have calculated $R(G,p)$ and
${\cal B}$, we find that this locus consists of arcs (and in some cases a line
segment) and does not enclose regions in the complex $p$ plane.  In some simple
cases ${\cal B}$ is connected, but in general, it may consist of several
disjoint components.  Again, this is quite different from what we found for the
loci ${\cal B}$ for chromatic polynomials of lattice strips; for those, a
sufficient (not necessary) condition that this locus encloses areas is that the
strip has periodic longitudinal boundary conditions \cite{w,wcyl,bcc}.

\item

In all of the cases for which we have calculated exact results, whenever
${\cal B}$ consists of an arc of a circle, then this circle is centered at
$p=1$.  More generally, even if ${\cal B}$ is not an arc of a circle, it is
roughly centered around $p=1$.  For families without multiple edges, we find
that the component(s) of ${\cal B}$ is (are) roughly concave toward the left.

\item 

The width $L_y \to \infty$ limit of loci ${\cal B}$ for the infinite-length
lattice strips is not necessarily the same as the continuous accumulation locus
of the zeros of the reliability polynomial for the two-dimensional
thermodynamic limit $L_x \to \infty$, $L_y \to \infty$ with $L_y/L_x$ fixed to
a finite nonzero constant.  This is clear since for any finite $L_y$ regardless
of how large, the infinite-length strip is a quasi-one-dimensional system,
characterized by the ratio $L_y/L_x = 0$.  Thus, for example, the Potts model
on these strips has a paramagnetic-ferromagnetic (PM-FM) phase transition point
only at $K=\infty$, i.e., $v=\infty$, while for 2D lattices, it has such a
PM-FM phase transition point at finite $v$.  However, in our previous studies
of continuous accumulation sets of loci of Potts model partition functions
(e.g., \cite{a}-\cite{sdg} and our earlier work referred to therein on
complex-temperature phase diagrams for Ising models) we have found that it is
possible to gain some insight into certain features of the continuous
accumulation set of zeros of the Potts model for the two-dimensional
thermodynamic limit from studies of wider and wider infinite-length strips.

In this context, we note that for the (infinite) square lattice, the PM-FM
phase transition point for the Potts model is given by $v^2=q$ \cite{wurev}.
This transition point is reflected in a nonanalyticity in the free energy of
the Potts model at the corresponding temperature variable $v$.  As we have
discussed and studied in earlier papers, e.g., \cite{chisq,chitri}, when one
generalizes $v$ to complex values, one sees that a singular point for a real
physical value of $v$, such as the PM-FM transition point, is associated with
the fact that a complex-temperature phase boundary crosses the real physical
$v$ axis at this point.  At the crossing points there are physical or
complex-temperature singularities in the free energy \cite{chisq}-\cite{p2}.
The boundaries on ${\cal B}$ (here as a function of $v$ for fixed $q$, but more
generally, also as a function of $q$ for fixed $v$) arise as the continuous
accumulation set of the complex-temperature zeros of the partition function,
analogous to the origin of the locus ${\cal B}$ studied here as continuous
accumulation set of the zeros of the reliability polynomial.  Connected with
this, the physical paramagnetic, ferromagnetic, and antiferromagnetic (AFM)
phases have complex-temperature extensions, and these are separated from each
other by the phase boundaries ${\cal B}$ in the $v$ plane.  We recall that for
the Potts model on the square lattice, the PM and FM phases occupy the
intervals $0 \le v \le \sqrt{q}$ and $\sqrt{q} \le v \le \infty$, respectively.
Analytically continuing $q$ to real values, using eq. (\ref{cluster}), we infer
that in the limit $q \to 0$ that yields the reliability polynomial, as given by
eq. (\ref{rz}), the PM phase therefore contracts to a point at $v=0$, which
corresponds, via the relation $p=v/(1+v)$ in eq. (\ref{pek}), to the point
$p=0$, while the FM phase expands to occupy the interval $0 \le v \le \infty$,
corresponding to the interval $0 \le p \le 1$.  As is evident, e.g., in the
square-lattice Ising model and more generally for the 2D $q$-state Potts model,
the complex-temperature extension of the FM phase includes part of the negative
real $v$ axis extending to $v=-\infty$ (as well as the outerlying region $|v|
\to \infty$ in the $v$ plane).  This semi-infinite interval of large negative
real values of $v$ is mapped, via the transformation $p=v/(1+v)$, to a finite
interval of values of $p \ge 1$.  Points in the outer region $v=Re^{i\theta}$
with $R >> 1$ are mapped via this transformation to points in the vicinity of
$p=1$, namely $p \simeq 1- R^{-1}e^{-i\theta}$.  Since (in the case $q \to 0$
relevant here) all of the points mentioned above in the $v$ plane, namely (i)
the interval $0 \le v \le \infty$, (ii) the interval of large negative real
$v$, and (iii) the semi-infinite annular region $v=Re^{i\theta}$ with $R \to
\infty$ are in the complex-temperature extension of the FM phase of the Potts
model, it follows that the images of these sets of points are all in the same
analytically connected region of the $p$ plane as defined by the locus ${\cal
B}$.  We can draw on another source of information about this boundary as
follows.  Arguments have been given that for the Potts model on the (infinite)
square lattice, the paramagnetic-antiferromagnetic (PM-AFM) transition occurs
at a (physical) root of the equation $v(v+4)+q=0$ \cite{baxter82}.  Setting
$q=0$ and $v=p/(1-p)$ as in (\ref{rz}) and (\ref{pek}), one finds the solutions
$v=0$ and $v=-4$, corresponding respectively to $p=0$, the solution already
found above, and $p=4/3$.  Hence for the square lattice, the image, under the
mapping $v \to p=v/(1+v)$, of the complex-temperature extension of the FM phase
includes the real interval $0 \le p \le 4/3$.  The complementary intervals
$-\infty \le p \le 0$ and $4/3 \le p \le \infty$ would be part of the image
under this map of an unphysical phase denoted the ``O'' (for ``other'') phase
in Ref. \cite{chisq} and our subsequent papers on complex-temperature phase
diagrams.  Using the fact that the complex-temperature phase boundary separates
these various phases, we infer that for the reliability polynomial on the
square lattice, in the thermodynamic limit, the zeros accumulate on a locus
${\cal B}$ that is a closed curve crossing the real axis at $p=0$ and $p=4/3$
and separating the interior region from the exterior, which latter includes the
intervals $-\infty \le p \le 0$ and $4/3 \le p \le \infty$.  In an obvious
nomenclature extending that of \cite{chisq}, we denote the regions in the
interior and exterior of this closed curve as $R_{FM}$ and $R_O$, indicating
the correspondence with the complex-temperature phases of the Potts model.
Just as the free energy is an analytic function in the interior of a
complex-temperature phase, the function $r$ that we have introduced in
\cite{ka3} and here in eq. (\ref{r}) is an analytic function in the regions
bounded by the locus ${\cal B}$.  One expects qualitatively similar results to
hold for the regions of analyticity of $r$ on (the thermodynamic limits of)
lattice graphs of dimensionality $d \ge 3$.  For the (infinite) square lattice,
\beq
p_c(sq) = \frac{4}{3}
\label{pcsq}
\eeq
We shall discuss how this compares with our exact results for $p_c$ values on
infinite-length finite-width strips below.  

Similar reasoning can be applied to obtain inferences for the loci ${\cal B}$
and the corresponding region diagrams of analyticity of the function $r$ on
other (infinite) 2D lattices.  For the triangular and honeycomb lattices, the
PM-FM phase transition point is given by a (physical) root of the equations
$v^2(v+3)-q=0$ and $v^3-3qv-q^2=0$, respectively \cite{kj}.  Evaluating these
for $q=0$ and $v=p/(1-p)$ yields for the triangular lattice, the solutions
$p=0$ and $p=3/2$, so that
\beq
p_c(tri)=\frac{3}{2}
\label{pctri}
\eeq
For the honeycomb lattice, the same evaluation yields just the solution $p=0$
and hence does not determine $p_c(hc)$ (By the reasoning given above, one can
infer that $p_c(hc) > 1$). The fact that the point $p=0$ on ${\cal B}$ is
common to all of the three lattices - square, triangular, and honeycomb -
follows because it is the image under the map (\ref{pek}) of the point $v=K=0$.
This point corresponds to infinite temperature, where, as one can see from
eqs. (\ref{zfun}) and (\ref{ham}), the spin-spin interaction does not
contribute to $Z(G,q,v)$, which reduces to $q^{|V|}$ for all of these lattices
(indeed for any graph $G$).

\item

For a given type of lattice strip (square, triangular, honeycomb, etc.) and a
given set of transverse boundary conditions, we may consider the sequence of
loci ${\cal B}$ for each width, $L_y$, and inquire whether these approach a
limiting locus as $L_y \to \infty$.  Our results are summarized in Table
\ref{properties} and are consistent with the hypothesis that such a limiting
locus exists.  As part of this analysis, we have investigated, for each type of
strip family, how $p_c$ or $(p_c)_{eff}$ depends on $L_y$.  We find that, for a
given type of lattice strip and transverse boundary conditions, if a value of
$p_c$ exists, i.e., if ${\cal B}$ crosses (intersects) the real axis, then
$p_c$ decreases as $L_y$ increases.  However, if, for a given type of lattice
strip and transverse boundary conditions, one considers the set of values of
both $p_c$ and $(p_c)_{eff}$ (the latter for widths where ${\cal B}$ comes
close to but does not cross the real axis), then the dependence on $L_y$ does
not appear to be monotonic.  For example, for the infinite-length limit of the
strip of the square lattice with free transverse boundary conditions (and any
longitudinal boundary conditions), for the width $L_y=2$, we have shown that
$p_c=4/3$.  Interestingly, this is the same value as inferred above for the
infinite square lattice defined via the usual thermodynamic limit.  For the
square-lattice strip with width $L_y=3$, the locus ${\cal B}$ does not cross
the real axis, but, as is evident in Fig. \ref{sqpxy3}, the inner endpoints of
the arcs on ${\cal B}$ are sufficiently close to the real axis that one can
infer a $(p_c)_{eff}$, and it is $(p_c)_{eff} \simeq 1.335$, which is slightly
greater than the value of $p_c$ for the $L_y=2$ strip and the inferred value of
$p_c(sq)$. As is evident in Table \ref{properties}, the values of $p_c$ are not
as close to $p_c(sq)$ for the square-lattice strips of the corresponding widths
with periodic transverse boundary conditions or self-dual boundary conditions,
as compared with free transverse boundary conditions, although they are again
consistent with approaching $p_c(sq)=4/3$ as the width increases.  For the
$L_y=2$ strip of the triangular lattice, we find the exact result $p_c=3/2$.
Again, interestingly, this is equal to the inferred value of $p_c(tri)$ for the
infinite triangular lattice defined via the thermodynamic limit.  We have
encountered this sort of situation before.  For example, for the chromatic
polynomial $P(G,q) = Z(G,q,v=-1)$ and its asymptotic limiting function as $|V|
\to \infty$, $W(\{G\},q)$ (the degeneracy per vertex for the Potts model), we
found that for the infinite-length strip of the triangular lattice with width
$L_y=3$ and toroidal or Klein-bottle boundary conditions, $q_c$ (defined as the
maximal point where the continuous accumulation set of zeros of $P(G,q)$
intersects the real $q$ axis) is $q_c=4$ \cite{tk}, the same value as for the
infinite triangular lattice.  Similarly, for infinite-length limits of
self-dual strips of the square lattice, we found $q_c=3$ \cite{dg,sdg}, the
same value as for the infinite square lattice.

As regards the behavior of ${\cal B}$ near $p=0$, our results are also
consistent with the possibility that for each type of lattice, as the strip
width $L_y \to \infty$, the respective loci ${\cal B}$ have, on the left,
complex-conjugate arc endpoints that come together and join at $p=0$, as
motivated by the discussion above for the infinite 2D lattices.  The observed
feature that this occurs only as a limit, in contrast to the crossing on the
right at the respective values of $p_c$, can be understood via the relation
(\ref{rz}) and the aforementioned fact that the infinite-length finite-width
strips are quasi-1D systems as far as their thermodynamic behavior is
concerned.  Since, as we are noted above, the interior of ${\cal B}$ for the
reliability polynomial on the 2D lattice corresponds to the image of the
complex-temperature extension of the FM phase of the 2D Potts model (for $q \to
0$), and since the argument that this must be separated from other phases
relies on the existence of nonzero ferromagnetic long range order
(magnetization), it follows that for quasi-1D systems, where for a short-range
interaction like that in (\ref{ham}), the standard Peierls argument shows that
there is no long range order, there is no FM phase.  This yields a plausible
explanation of the features that we observe in our exact solutions for the
infinite-length, finite-width strip graphs, that the loci ${\cal B}$ do not
enclose regions in the $p$ plane.

\end{itemize}

\begin{table}
\caption{\footnotesize{Properties of reliability polynomials and resultant loci
${\cal B}$ including $p_c$ or $(p_c)_{eff}$ for various lattice strip
graphs. The notation $sqdl$ and $D$ refer to the self-dual strips of the square
lattice.}}
\begin{center}
\begin{tabular}{|c|c|c|c|c|c|c|}
\hline\hline $G_s$ & $L_y$ & $BC_y$ & $BC_x$ & $d_{eff}$ & $N_\alpha$ &
$p_c$ \\ \hline\hline
sq   & 1 & F &    F & 2   & 1    & $-$     \\ \hline
sq   & 1 & F &    P & 2   & 1    & $-$     \\ \hline
sq   & 2 & F &    F & 3   & 2    & 4/3     \\ \hline
sq   & 2 & F & (T)P & 3   & 3    & 4/3     \\ \hline
sq   & 3 & F &    F &10/3 & 4    & 1.335   \\ \hline
sq   & 3 & F & (T)P &10/3 & 4    & 1.335   \\ \hline\hline
sq   & 2 & P &    F & 4   & 2    &  2      \\ \hline
sq   & 2 & P & (T)P & 4   & 3    &  2      \\ \hline
sq   & 3 & P &    F & 4   & 3    & 1.402   \\ \hline
sq   & 3 & P & (T)P & 4   & 11   & 1.402   \\ \hline
sq   & 4 & P &    F & 4   & 6    & 1.384   \\ \hline
sqdl & 1 & D &    F & 4   & 2    & 3/2     \\ \hline
sqdl & 1 & D &    P & 4   & 3    & 3/2     \\ \hline
sqdl & 2 & D &    F & 4   & 5    & 1.4     \\ \hline
sqdl & 2 & D &    P & 4   & 10   & 1.4     \\ \hline
sqdl & 3 & D &    F & 4   & 14   & 1.386   \\ \hline
sqdl & 3 & D &    P & 4   & 35   & 1.386   \\ \hline\hline
tri  & 2 & F &    F & 4   & 2    & 3/2    \\ \hline
tri  & 2 & F & (T)P & 4   & 3    & 3/2    \\ \hline\hline
hc   & 2 & F &    F & 5/2 & 2    & 6/5    \\ \hline
hc   & 2 & F & (T)P & 5/2 & 3    & 6/5    \\ \hline\hline
\end{tabular}
\end{center}
\label{properties}
\end{table}

\section{Conclusions}

In this paper we have presented exact calculations of reliability polynomials
$R(G,p)$ for lattice strips $G$ of fixed width and arbitrarily great length
with various boundary conditions.  We have introduced the notion of a
reliability per vertex, $r(\{G\},p)$ and have calculated this exactly for the
infinite-length limits of various lattice strip graphs.  We have also studied
the zeros of $R(G,p)$ in the complex $p$ plane and determine exactly the
asymptotic accumulation set of these zeros ${\cal B}$, across which $r(\{G\})$
is nonanalytic.  We have observed and discussed several general features of the
$r$ functions and the loci ${\cal B}$ for these families of graphs.  

Acknowledgment: This research was partially supported by the NSF grant
PHY-00-98527.

\vfill
\eject
\end{document}